\documentclass[twocolumn]{aastex631}
\usepackage{gensymb}
\usepackage{xspace}
\usepackage{booktabs}
\usepackage[flushleft]{threeparttable}
\usepackage{hyperref}
\usepackage{amsmath}
\usepackage{listings}
\newcommand{\keckhires}{Keck-HIRES\xspace}

\newcommand{\radvel}{\texttt{radvel}\xspace}
\newcommand{\jwst}{\textit{JWST}\xspace} 
\newcommand{\gaia}{\textit{Gaia}\xspace}
\newcommand{\mearth}{$M_\mathrm{\oplus}$\xspace}

\newcommand{\mjup}{$M_\mathrm{Jup}$\xspace}
\newcommand{\teff}{$T_\mathrm{eff}$\xspace}
\newcommand{\mpsini}{$m_\mathrm{p} \sin{i}$\xspace}

\shorttitle{Direct imaging and RV of Wolf 359}
\shortauthors{Bowens-Rubin et al.}

\graphicspath{{./}{figures/}}

\begin{document}

\title{A Wolf 359 in sheep's clothing: Hunting for substellar companions in the fifth-closest system using combined high-contrast imaging and radial velocity analysis}

\correspondingauthor{Rachel Bowens-Rubin}
\email{rbowru@ucsc.edu}

\author[0000-0001-5831-9530]
{Rachel Bowens-Rubin}
\affiliation{Astronomy Department, University of California Santa Cruz,
1156 High St, Santa Cruz, CA 95064, USA}

\author[0000-0001-8898-8284]{Joseph M. Akana Murphy}
\altaffiliation{NSF Graduate Research Fellow}
\affiliation{Astronomy Department, University of California Santa Cruz,
1156 High St, Santa Cruz, CA 95064, USA}

\author[0000-0002-1954-4564,]{Philip M. Hinz}
\affiliation{Astronomy Department, University of California Santa Cruz,
1156 High St, Santa Cruz, CA 95064, USA}

\author[0000-0002-9521-9798]{Mary Anne Limbach}
\affiliation{Department of Astronomy, University of Michigan, Ann Arbor, MI 48109, USA}

\author[0000-0003-4526-3747]{Andreas Seifahrt}
\affiliation{Department of Astronomy \& Astrophysics, University of Chicago, 5640 South Ellis Avenue, Chicago, IL
60637, USA}

\author[0000-0003-2102-3159]{Rocio Kiman}
\affiliation{Department of Astronomy, California Institute of Technology, Pasadena, CA 91125, USA}
\affiliation{Kavli Institute for Theoretical Physics, University of California Santa Barbra,
Kohn Hall, Santa Barbara, CA 93106}

\author[0000-0002-5082-6332]{Ma\"issa Salama}
\affiliation{Astronomy Department, University of California Santa Cruz,
1156 High St, Santa Cruz, CA 95064, USA}

\author[0000-0003-1622-1302]{Sagnick Mukherjee}
\affiliation{Astronomy Department, University of California Santa Cruz,
1156 High St, Santa Cruz, CA 95064, USA}

\author[0000-0003-2404-2427]{Madison Brady}
\affiliation{Department of Astronomy \& Astrophysics, University of Chicago, 5640 South Ellis Avenue, Chicago, IL
60637, USA}

\author[0000-0001-5365-4815]{Aarynn L. Carter}
\affiliation{Astronomy Department, University of California Santa Cruz,
1156 High St, Santa Cruz, CA 95064, USA}

\author[0000-0003-0054-2953]{Rebecca Jensen-Clem}
\affiliation{Astronomy Department, University of California Santa Cruz,
1156 High St, Santa Cruz, CA 95064, USA}

\author[0000-0002-9805-3666]{Maaike A.M. van Kooten}
\affiliation{Herzberg Astronomy and Astrophysics Research Centre,
5071 West Saanich Rd, Victoria, BC, V9E 2E7, Canada}

\author[0000-0002-0531-1073]{Howard Isaacson}
\affiliation{Department of Astronomy, University of California, Berkeley, CA 94720, USA}
\affiliation{Centre for Astrophysics, University of Southern Queensland, Toowoomba, QLD, Australia}

\author[0000-0002-6115-4359]{Molly Kosiarek}
\altaffiliation{NSF Graduate Research Fellow}
\affiliation{Astronomy Department, University of California Santa Cruz,
1156 High St, Santa Cruz, CA 95064, USA}

\author[0000-0003-4733-6532]{Jacob L.\ Bean}
\affiliation{Department of Astronomy \& Astrophysics, University of Chicago, 5640 South Ellis Avenue, Chicago, IL
60637, USA}

\author[0000-0003-0534-6388]{David Kasper}
\affiliation{Department of Astronomy \& Astrophysics, University of Chicago, 5640 South Ellis Avenue, Chicago, IL
60637, USA}

\author[0000-0002-4671-2957]{Rafael Luque}
\affiliation{Department of Astronomy \& Astrophysics, University of Chicago, 5640 South Ellis Avenue, Chicago, IL
60637, USA}

\author[0000-0001-7409-5688]{Gudmundur Stef\'ansson}
\affiliation{Department of Astrophysical Sciences, Princeton University, 4 Ivy Lane, Princeton, NJ 08540, USA}
\altaffiliation{NASA Sagan Fellow}

\author[0000-0002-4410-4712]{Julian St{\"u}rmer}
\affiliation{Landessternwarte, Zentrum f{\"u}r Astronomie der Universität Heidelberg, K{\"o}nigstuhl 12, D-69117 Heidelberg, Germany}

\begin{abstract}
Wolf 359 (\textit{CN Leo, GJ 406, \gaia DR3 3864972938605115520}) is a low-mass star in the fifth-closest neighboring system (2.41\,pc). Because of its relative youth and proximity, Wolf 359 offers a unique opportunity to study substellar companions around M stars using infrared high-contrast imaging and radial velocity monitoring. We present the results of Ms-band (4.67\,$\mu$m) vector vortex coronagraphic imaging using Keck-NIRC2 and add 12 Keck-HIRES velocities and 68 MAROON-X velocities to the radial velocity baseline.  
Our analysis incorporates these data alongside literature radial velocities from CARMENES, HARPS, and Keck-HIRES to rule out the existence of a close ($a < 10$\,AU) stellar or brown dwarf companion and the majority of large gas-giant companions. Our survey does not refute or confirm the long-period radial velocity candidate, Wolf 359\,b ($P\sim2900$\,d) but rules out the candidate's existence as a large gas-giant ($>4$\mjup) assuming an age of younger than 1\,Gyr. We discuss the performance of our high-contrast imaging survey to aid future observers using Keck-NIRC2 in conjunction with the vortex coronagraph in the Ms-band and conclude by exploring the direct imaging capabilities with \jwst to observe Jupiter-mass and Neptune-mass planets around Wolf 359.

\end{abstract}


\section{Introduction\label{sec:intro}}

Over 70\% of the stars in our galaxy are M-dwarfs, yet we know little about the exoplanets that exist in these systems beyond the snow line ($\gtrsim$0.5 AU, \citealt{Mulders2015}). Most exoplanet detection methods and surveys are blind to this discovery space. 
The geometric probability of an exoplanet transit occurring for an exoplanet orbiting an M-dwarf beyond 1\,AU is less than $0.1$\%. 
Astrometry and radial velocity surveys of M-dwarfs require lengthy baselines in order to observe a planet's full orbit because planets orbiting low-mass stars have longer periods for an equivalent separation. 

Microlensing surveys have provided the first hint that cold gas giants, ice giants, and super-Earths could be common outside the snow line of M-dwarfs with increasing prevalence for smaller planets. A survey from \cite{Cassan2012} estimated that the majority of low-mass stars host a giant planet between 0.5--10 AU, with Jupiter-like planets ($0.3-10$ \mjup) at an occurrence rate of $17^{+6}_{-9}\%$, Neptune-like planets ($10-30$ \mearth) with a rate of $52^{+22}_{-29}\%$, and super-Earths ($5-10$ \mearth) with a rate of $62^{+35}_{-37}\%$. A microlensing survey by the Microlensing Observations in Astrophysics collaboration is consistent with these results and concluded that Neptune-sized planets are one of the most common types of planet seen outside the snow line \citep{Suzuki2016}. \citealt{Poleski2021} used data from the Optical Gravitational Lensing Experiment to determine that nearly every star could host an ice-giant planet from 5-15\,AU, measuring an occurrence rate of $1.4^{+0.9}_{-0.6}$ ice giants per system. 

Exoplanet direct imaging---where photons from an exoplanet are spatially resolved from their host star---is the only exoplanet detection technique that offers a pathway for characterizing the atmosphere, composition, and formation history for exoplanets orbiting beyond the snow line that are unlikely to transit. When directly imaging the closest set of stellar neighbors ($d < 5$ pc), the current generation of high-contrast imaging systems on 8--10 m telescopes can probe comparatively colder planets at angular separations corresponding to where the prevalence of exoplanets outside the snow line is expected to peak (1--10\,AU; \citealt{Fernandes2019}).
Proximity in stellar distance makes companions appear at proportionally wider separation angles from their host star for a given orbit ($\theta_\mathrm{sep} \propto a/d$) and boosts the apparent magnitude of the companion logarithmically ($m = 5 \log_{10}(d/10$ pc$) + M$). This makes companions that are dimmer in absolute magnitude and closer in orbital separation easier to detect than if they were in a more distant analogous system. 

The heritage of detecting exoplanets via the direct imaging technique has been to conduct blind surveys of hundreds of young-star systems in search of a rare set of large gas giant planets on long-period orbits that are bright enough to detect using short integration times.
Thanks to the growing abundance of long-baseline exoplanet radial velocity (RV) data \citep[e.g.,][]{rosenthal21, Trifonov2020, Ribas2023}, 
we can now use RV data in tandem with high-contrast imaging (HCI) observations to tailor our imaging observations to conduct lengthier measurements around fewer systems. 
Information from RV data can be applied to select viable targets for imaging, choose the optimal imaging filters, predict how much integration time is required, and predict when a companion will be at its maximum separation from its host star.
This targeted approach to HCI observing motivates the use of extended observing sequences which can expand our abilities to directly image colder ($<500$\,K) companions.

In many cases, we only need a hint to a companion's existence to curate an HCI observation using RV data. \cite{Cheetham2018} demonstrated this by leveraging RV data to directly image an ultra-cool brown dwarf, HD 4113C. Based on the the CORALIE survey's detection of long-term RV trends \citep{Udry2000}, \cite{Rickman2019} conducted targeted direct imaging resulting in the discovery of three giant planets and two brown dwarfs. The TRENDS high-contrast imaging survey used long-baseline velocities from \keckhires to target their survey for white dwarf and substellar companions (e.g., \citealt{Crepp2018}, \citealt{Crepp2016}). 
\cite{Hinkley2022} used the VLTI/GRAVITY instrument to discover HD 206893\,c by utilizing long-baseline RV data from European Southern Observatory's High Accuracy Radial velocity Planet Searcher (HARPS, \citealt{Pepe2002}; \citealt{Mayor2003}) and correlating it with the \gaia-Hipparcos astrometry accelerations (\citealt{HGCA}) and orbital astrometry of the system's outer companion. 

Conducting targeted HCI observations of nearby systems that span multiple nights is becoming an increasingly common observing strategy to probe for sub-Jupiter mass exoplanets. The surveys from \citealt{Mawet2019} and \citealt{Llop-Sayson2021} %used this to their advantage to 
completed multi-night HCI campaigns of the nearby, youthful $\varepsilon$ Eridani system ($d = 3.22$ pc, age = $600\pm200$ Myr) with the goal of directly detecting the RV-discovered exoplanet, $\varepsilon$ Eridani b.  Combined, the 2017 and 2019 surveys collected nearly 16 hours of 4.67$\mu$m imaging data over nine nights using the W. M. Keck Observatory's NIRC2 Imager \citep[Keck-NIRC2;][]{wizinowich00} but were not able to make an imaging detection of the planet. By combining the mass upper-limits from HCI with RV and \gaia accelerations, \cite{Llop-Sayson2021} constrained the mass of $\varepsilon$ Eridani b to be in the sub-Jupiter mass domain, $0.66^{+0.12}_{-0.09}$ \mjup.  
\cite{Wagner2021} also demonstrated the advantage of searching for companions around nearby stars by performing a 100 hr HCI survey at 10--12.5 $\mu$m of the $\alpha$ Centauri system ($d = 1.3$ pc, age $= 5.3 \pm 0.3$\,Gyr). They imaged one candidate and demonstrated that it was possible to achieve survey sensitivities down to warm sub-Neptune mass planets through the majority of the $\alpha$ Centari habitable zone. While these surveys were not able to make definitive direct detections, they demonstrated the possibilities  of future ground-based mid-infrared HCI campaigns of nearby stars.

In this paper, we present the results of our joint HCI-RV survey to search for substellar companions around the solar-neighborhood star Wolf 359 (\textit{CN Leo, GJ 406, \gaia DR3 3864972938605115520}). 
The paper is organized as follows: 
In the remainder of \S\ref{sec:intro}, we provide an overview of the Wolf 359 system.
In \S\ref{sec:obs}, we report our observational and data reduction methods for the Keck-NIRC2 coronographic imaging survey and the RV measurements from the W. M. Keck Observatory High Resolution Echelle Spectrometer \citep[\keckhires;][]{HIRES} and Gemini-North MAROON-X spectrograph \citep{MAROONX}.
In \S\ref{sec:analysis}, we estimate Wolf 359's stellar age, apply these age constraints to the HCI data to set companion mass upper bounds, and provide an updated RV analysis combining our measurements with the previously published RV data from HARPS, \keckhires, and CARMENES. 
In \S\ref{sec:discussion}, we discuss how our imaging performance with Keck-NIRC2 compared to the predicted performance and then explore what JWST high-contrast imaging could reveal about the Wolf 359 system.

\subsection{The Wolf 359 System} \label{subsec:introwolf359}
Wolf 359 is a solar-metallicty M6V star \citep{Pineda2021} and one of our nearest stellar neighbors\footnote{As one of our nearest neighbors, this system has captured the public's interest and is a setting in many fictional stories including the \href{https://wolf359.fm/}{Wolf 359 podcast} and several episodes in the \textit{Star Trek} franchise.} \citep[2.41 pc;][]{GaiaDR3}. Table \ref{tab:PropofWolf359} summarizes Wolf 359's stellar parameters. 

Radial velocity surveys have been monitoring Wolf 359 for more than two decades. 
A preprint paper presented by \cite{Tuomi2019} identified two exoplanet candidates orbiting Wolf 359 using 63 RV measurements from \keckhires and HARPS  spanning 13 years. These planet candidates are summarized in Table \ref{Tab:ExoplanetCandidates}. The shorter-period candidate (Wolf 359\,c) was refuted by \cite{Lafarga2021} after determining that the RV signal matched the star's rotation period. The RV signal for the Wolf 359\,b candidate could correspond to a cold, Neptune-like exoplanet on a wide orbit of approximately 8 years \citep[$P_{orb} = 2938 \pm 436$\,d,  $a = 1.845^{+0.289}_{-0.258}$\,AU;][]{Tuomi2019}.  

Wolf 359 has conflicting age estimates in the literature, but most indicate that the star is young ($<1$\,Gyr).  The star is highly active, with stellar flares that occur approximately once every 2\,hr \citep{Lin2022}.  Wolf 359 has strong flare activity even among flaring M dwarfs \citep{Lin2021}, which is consistent with a youthful age estimate.   
An age estimate by \citealt{Pavlenko2006} made by modeling the spectral energy distribution predicts that Wolf 359 could be as young as $100-350$\,Myr, which is consistent with its high activity. 
Wolf 359 also has a fast rotation period \citep[$P_\mathrm{rot} = 2.705 \pm 0.007$ d;][]{Guinan2018wolf359rot}, as confirmed with photometry from \textit{K2} \citep{howell14}, among other observatories.
The combination of the gyrochronological relationship from \cite{Engle2018} and Wolf 359's stellar rotation period suggests an age estimate of $<500$ Myr. However, the star lies at the edge of \cite{Engle2018}'s rotation–activity–age relationship for M0-6 stars, and the rotation period cannot act as a direct proxy for age in this system in this context.   

The combination of Wolf 359's proximity and potential youth make it an ideal system for searching for companions using infrared direct imaging. 
An exoplanet candidate like Wolf 359\,b would not be possible to directly image around most star systems. However, because Wolf 359 is one of our nearest neighbors, the parameters of the Wolf 359\,b candidate can be constrained 
using our current generation of HCI instruments operating at 8-10\,m telescopes.

\begin{deluxetable}{c c}
\tablecaption{Properties of Wolf 359 \label{tab:PropofWolf359}  }
%\tabletypesize{\scriptsize}
%\begin{tabular}{cc}
\tablehead{\colhead{Property} & \colhead{Value}  }
\startdata
RA J2000        & 10 56 28.92 $^{(a)}$                  \\
DEC J2000       & +07 00 53.00 $^{(a)}$                 \\
Distance        & $2.4086 \pm 0.0004$\,pc $^{(a)}$       \\
Parallax        & $415.18 \pm 0.07$\,mas $^{(a)}$        \\
Spectral Type   & solar-metallicity M6 $^{(b)}$         \\
Mass            & $0.110 \pm 0.003~M_{\odot}$ $^{(c)}$  \\
Teff            & $2749^{+44}_{-41}$\,K $^{(c)}$         \\
Radius          & $0.144 \pm 0.004~M_{\odot}$ $^{(c)}$  \\
log(g)          & $5.5$\,cgs $^{(d)}$                      \\
V mag           & 13.5 $^{(e)}$                         \\
R mag           & 11.684 $^{(e)}$                       \\
H mag           & 6.482 $^{(f)}$                        \\
MKO Ms mag      & $5.85 \pm 0.06$ $^{(g)}$ 
\\
Rotation Period & $2.705 \pm 0.007$\,d $^{(h)}$          \\
Age range       & 100\,Myr--1.5\,Gyr $^{(f)}$  
\enddata
\tablenotetext{}{\\
(a)	\citealt{GaiaDR3};	\\
(b)	\citealt{Kesseli2019};	\\
(c)	\citealt{Pineda2021};	\\
(d)	\citealt{Fuhrmeister2005};	\\
(e)	\citealt{Landolt2009};	\\
(f)	\citealt{Cutri2003};	\\
(g)	\citealt{Leggett2010};	\\
(h)	\citealt{Guinan2018wolf359rot};	\\
(f)	The lower estimate is from \citealt{Pavlenko2006} and the upper estimate is from the kinematic age estimated in Section \ref{subsec:age} of this work.	
}
\end{deluxetable}

\begin{deluxetable*}{c c c c c c}
\tablecaption{Exoplanet Candidates Identified by \cite{Tuomi2019} \label{Tab:ExoplanetCandidates}
}
%\tabletypesize{\largesize}
\tablehead{\colhead{Candidate} & \colhead{Period (d)} & \colhead{$m\sin{i} (M_{\bigoplus})$} & \colhead{a (AU)} & \colhead{Status}  & \colhead{Note}    } 
\startdata
Wolf 359 b         & $2938\pm436$                 & $43.9^{+29.5}_{-23.9}$              & $1.845^{+0.289}_{-0.258}$ & \begin{tabular}[c]{@{}c@{}}possible \\ cold-Neptune\end{tabular} & \begin{tabular}[c]{@{}c@{}}investigated in \\ this work\end{tabular}      \\
Wolf 359 c         & $2.6869^{+0.0004}_{-0.0003}$ & $3.8^{+2.0}_{-1.6}$                 & $0.018\pm0.002$           & false positive$^\dagger$                                                   & \begin{tabular}[c]{@{}c@{}}RV signal is due \\ to star rotation\end{tabular} 
\enddata
\tablenotetext{}{
$^\dagger$ Wolf 359 c was refuted by \citealt{Lafarga2021}.}
\end{deluxetable*}

%%%%%%%%%%%%%%%%%%%%%%%%%%%%%%%%%%%
\section{Observations and Data Reduction} \label{sec:obs}

\subsection{Keck-II NIRC2 Vortex Coronagraphy} \label{subsec:obsHCI}
We conducted high-contrast imaging observations of the Wolf 359 system with the W.M. Keck Observatory NIRC2 imager coupled with the vector vortex coronagraph \citep{Vortex}.  We completed our observations over three nights, as summarized in Table \ref{tab:HCI}. 

We conducted HCI observations using the fixedhex pupil stop with Keck's L/M-band vortex coronagraph. The telescope was operated in the vertical angle rotation mode (Sky PA $= 4.43^{\circ}$) to enable angular differential imaging (ADI) analysis methods. The centering of the vortex was controlled using the in-house QACTIS IDL software package \citep{QACITS}. Each QACITS sequence consisted of a set of 
(1) three calibration images to acquire an off-axis star PSF and sky images, (2) three optimization images to center the star on the vortex and stabilize the tip/tilt in the adaptive optics system, (3) a series of science images.    

We operated the Keck-II adaptive optics system with the recently commissioned near-infrared pyramid wavefront sensor (PyWFS) \citep{Bond2020} in natural guide star mode.  We selected the PyWFS over the facility Shack–Hartmann wavefront sensor because it is better suited for performing adaptive optics corrections when using an M-dwarf as a natural guide star because it operates in H-band (1.633$\mu m$, \href{https://www2.keck.hawaii.edu/inst/nirc2/filters.html}{NIRC2 Filters}) rather than 
R-band (0.641 $\mu m$, \citealt{Bessell2005}).  Wolf 359 is 5.2 magnitudes brighter in the H-band versus R-band \citep{Landolt2009, Cutri2003}, thus we were able to take advantage of the improved AO quality with the significantly more flux available for wavefront correction.  

Our HCI survey spanned three nights in 2021: February 22, February 23, and March 31 (UT).  We collected images using the Ms filter (4.670$\mu m$, \href{https://www2.keck.hawaii.edu/inst/nirc2/filters.html}{NIRC2 Filters}) with NIRC2 operated in narrow  mode. The science images had a frame size of of 512 x 512 pixels (5.090\arcsec\,x\,5.090\arcsec; pixel scale = 0.009942 $\pm$ 0.00005 arcsec/pixel, \href{https://www2.keck.hawaii.edu/inst/nirc2/genspecs.html}{Keck General Specs}). The frames were taken with an integration time of 0.3\,s with 90\,coadds. 
We obtained a total of 664 science frames over 14 QACTIS sequences, totaling 4.98\,hr of science integration time.

We performed our data reduction using the \emph{VIP: Vortex Imaging Processing} python package (\texttt{VIP}) \citep{VIP}.
We pre-processed the NIRC2 data for bad pixels, flat-field correction, and sky background correction using the automated pipeline described in \citealt{Xuan2018} using \texttt{VIP} version 0.9.9.  Sky subtraction was completed using the PCA-based approached described in \citealt{Hunziker2018} using \texttt{VIP} version 1.3.0. 
After pre-processing the science images, we removed 5\% of the lowest-quality science frames using \texttt{VIP}'s Pearson correlation bad-frame detection from each night. 

To establish an anchor for our reported contrast, we measured the flux of Wolf 359 using the unobstructed PSF images taken at the start of the QACTIS sequence. We created a PSF template by combining and then normalizing the 14 PSF images taken on 2021 March 31 (UT). The PSF frames were collected using an integration time per coadd of 0.015s with 100 coadds. We performed the stellar photometry using the \texttt{fit\_2dgaussian} function, as outlined in the \texttt{VIP} tutorial.  We measured the full width half max of the NIRC2 Ms PSF to be $FWHM = 9.67$\,pixel (0.0962\arcsec).

\begin{figure*}
\gridline{\fig{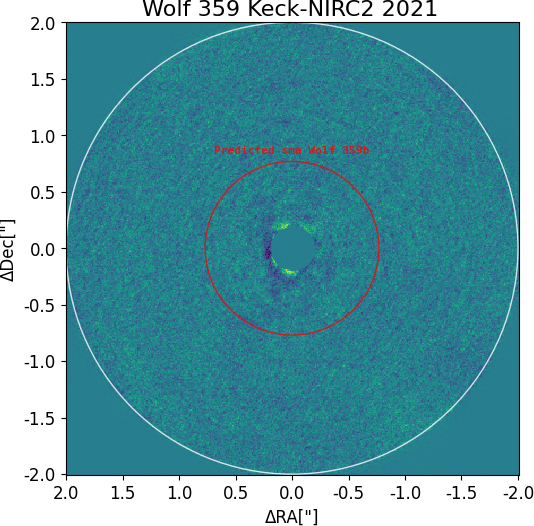}{0.42\textwidth}{(a) The final reduced image }
          \fig{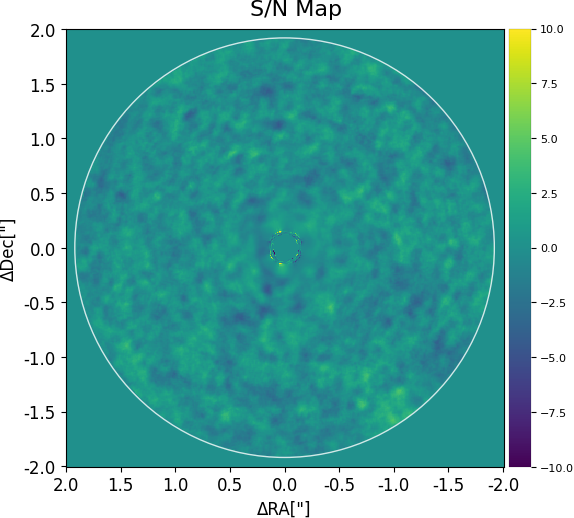}{0.46\textwidth}{(b) SNR Map}
          }
\caption{Final reduced image of the Wolf 359 system from the Keck-NIRC2 high-contrast imaging survey: Our final reduced image of the Wolf 359 system was created using the the highpass-filtered three-night combined image cube. The corresponding S/N map is shown in (b).  The red circle shown in (a) corresponds to the predicted semi-major axis of the Wolf 359\,b candidate.  The stellar PSF was subtracted using full-frame PCA with \texttt{VIP}. No companion-like point sources were detected to more than $2\sigma$ above the background using \texttt{VIP}'s built-in detection function.
\label{fig:HCIimg}}
\end{figure*}

We created the final reduced image using the combined image set from the three nights with the 631 images that passed bad-frame detection. 
We applied a highpass filter to each individual image using a \texttt{VIP}'s Gaussian highpass filter with size 2.25 $FWHM$. 
The images were then derotated using the parallactic angle and median combined. We subtracted the stellar point spread function (PSF) using full-frame angular differential imaging principle component analysis (PCA) using \texttt{VIP}'s \texttt{pca} module (following the methods of \citealt{Soummer2012} and \citealt{Amara2012}). We performed PCA optimization by injecting a fake companion 100\,pixels from the star to  determine the number of principle components that yielded the max signal-to-noise of the fake companion.   
The three-night combined image set had an optimal number of principal components of PC = 18 (PC = 4 when highpass filtering was applied). While performing PCA stellar point spread subtraction, we adopted a center masking of 2 $FWHM$ and a parallactic exclusion angle the size of 1 $FWHM$.   
The final reduced image from the highpass-filtered three-night combined image set is shown in Figure \ref{fig:HCIimg} along with its accompanying  signal-to-noise threshold map.  We detected no point source signals above a 2$\sigma$ threshold using VIP's built in detection function in log mode. We thus conclude that we did not detect any companions in the direct imaging portion of this survey.

\begin{figure*}
\gridline{\fig{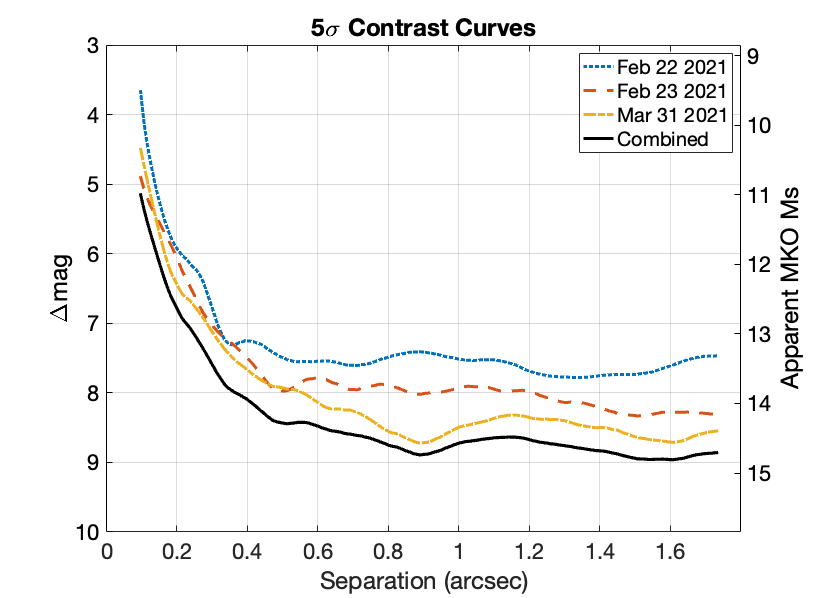}{0.5\textwidth}{(a) 5$\sigma$ contrast curves by observing night with the combined-image contrast}
\fig{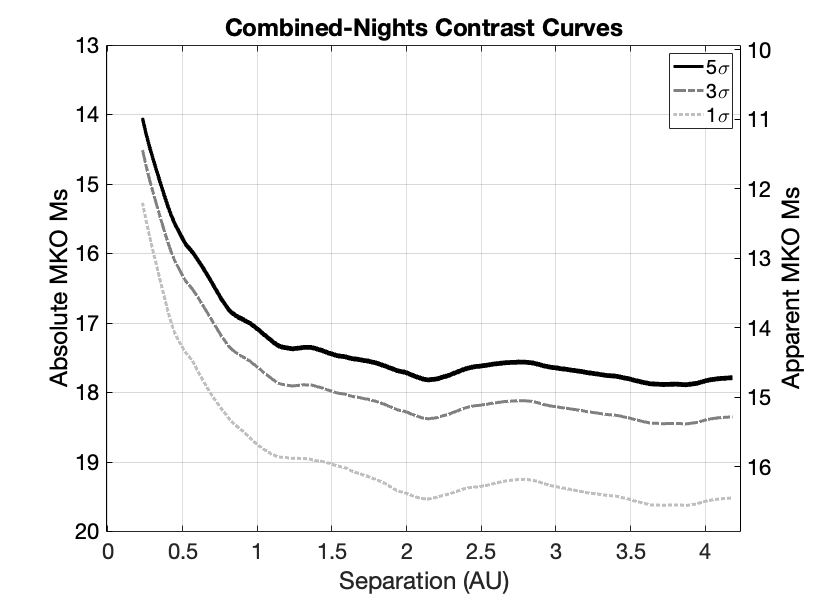}{0.5\textwidth}{(b) The planet sensitivity of the 3-night combined survey }}
\caption{Contrast curves from the Keck-NIRC2 imaging survey: The contrast curves were created using the fullframe PCA algorithm in \texttt{VIP} with the images that were not highpass filtered. The solid black line represents the 5$\sigma$ sensitivity achieved with the combined-nights cube.
\label{fig:cc}}
\end{figure*}

We calculated contrast curves using \texttt{VIP}'s \texttt{contrast\_curve} function, which calculates the $\sigma*noise/throughput$ using fake planet injection with a student-t distribution correction.  We found that applying a highpass filter had little affect on our final sensitivity, so our contrast curves are reported using the images with no applied highpass filtering.   The combined-night contrast curve was calculated by first processing the sensitivity by separation for each night separately. The combined-nights sensitivity was then calculated using a weighted variance at each separation, $\sigma_{comb}(sep) = \sqrt{1/(\sigma_{n1}(sep)^{-2}+\sigma_{n2}(sep)^{-2}+\sigma_{n3}(sep)^{-2}})$. The overall sensitivity of the HCI survey of Wolf 359 is plotted in Figure \ref{fig:cc}.

\begin{deluxetable*}{ccccccccc}
\tablecaption{High-Contrast Imaging Keck-NIRC2 Observing Summary \label{tab:HCI}}
%\tabletypesize{\largesize}
\tablehead{\colhead{Date (UT)} 
&  \colhead{\begin{tabular}[c]{@{}c@{}}Total Sci \\ Frames\end{tabular}} 
& \colhead{\begin{tabular}[c]{@{}c@{}}Total Int \\ Time (hr)\end{tabular}}
&  \colhead{PA Change}
&  \colhead{\begin{tabular}[c]{@{}c@{}}PWV\\ opacity \end{tabular}}
& \colhead{\begin{tabular}[c]{@{}c@{}}Optimal PC \\ full fr. PCA\end{tabular}}
& \colhead{\begin{tabular}[c]{@{}c@{}}$5\sigma \Delta Mag$\\ 0.2'' (0.5 AU)\end{tabular} }
& \colhead{\begin{tabular}[c]{@{}c@{}}$5\sigma \Delta Mag$\\ 0.98'' (1 AU)\end{tabular}}
& \colhead{\begin{tabular}[c]{@{}c@{}}$5\sigma \Delta Mag$\\ 1.7'' (4 AU)\end{tabular}}
} 
\startdata
2021 Feb 22       & 181  & 1.36  & 77.35$^{\circ}$  & 0.08-0.20  & 8  & 6.0   & 7.3 & 7.5    \\
2021 Feb 23       & 200 & 1.50   & 76.88$^{\circ}$ &  0.06-0.10 & 28  & 6.1 & 7.5 & 8.3  \\
2021 Mar 31       & 283  & 2.12  & 126.29$^{\circ}$ &  0.10-0.16  & 14  & 6.4  & 7.7 & 8.6 \\ \hline
Combined Nights    & 664  & 4.98 & -- & -- & 18 & 6.8  & 8.2  & 8.9     
\enddata
\tablenotetext{}{All science images were collected using the MKO Ms filter, $tint = 0.3s$, $coadd=90$, and $subframe size=512\times512$ with NIRC2 in narrow mode. The precipitable  water vapor opacity was measured at 225\,GHz by the Submillimeter Array and retrieved at \url{http://www.eao.hawaii.edu/weather/opacity/mk/archive/}.  The optimal PC and $5\sigma$ contrast is reported for the image sets with no highpass filter applied.
To convert the listed $\Delta Mag$ to MKO Ms apparent magnitude, add 5.85. }
\end{deluxetable*}

%%%%%%%%%%%%%%%%%%%%%%%%%%%%%%%
\subsection{Radial Velocity Observations} \label{subsec:obsRV}
\subsubsection{\keckhires}
We present an additional 12 \keckhires high-precision RV measurements gathered by the California Planet Search (CPS) team between Dec 25 2017 and Jan 13 2022 (UT). The \keckhires velocities are available in Appendix \ref{sec:RVappend} and 
online in a machine readable format.  These measurements extend the baseline of the \keckhires post-2004 velocities to over 17 years when combined with the 40 \keckhires RVs included in \cite{Tuomi2019}. The new \keckhires exposures were collected with the C2 decker (14\arcsec$\times$0\arcsec.86, $R =$ 45,000) and had a median integration time of 1800\,s, corresponding to a median SNR of 65 pix$^{-1}$ at 5500 \AA. 

Observations were taken with a warm ($50\degree$ C) cell of molecular iodine at the entrance slit \citep{butler96} and RVs were determined following the procedures of \cite{howard10}. The superposition of the iodine absorption lines on the stellar spectrum provides both a fiducial wavelength solution and a precise, observation-specific characterization of the instrument's PSF. Each RV spectrum was then modeled as the product of the deconvolved template spectrum and the FTS molecular iodine spectrum which is convolved with the point-spread function. The chi-squared value of this model is minimized with the RV ($Z$) as one of the free parameters.

RVs computed via the iodine cell method require a high-SNR iodine-free ``template'' of the stellar spectrum. Ideally, CPS aims for template spectra to have an SNR of about 200 pix$^{-1}$ at 5500 \AA\ in order to properly deconvolve the spectrum with the instrument's PSF, which is measured by observing rapidly rotating B stars immediately before and after the template exposure(s). In the case of Wolf 359, CPS acquired three consecutive iodine-free exposures of the star on 2005 Feb 27 with the B1 decker (3.5\arcsec$\times$0\arcsec.574, $R =$ 60,000). Each observation had an exposure time of 400 s corresponding to a combined SNR of 40 pix$^{-1}$ at 5500 \AA. Because Wolf 359 is relatively faint in $V$-band \citep[$V = 13.5$ mag;][]{Landolt2009}, high SNR \keckhires exposures quickly become prohibitively expensive (SNR of $\sim100$ pix$^{-1}$ would take well over an hour of integration). Rather than attempt to acquire another, higher SNR template of Wolf 359, we searched for a best-match template from a library of over 300 stars with high-SNR, iodine-free \keckhires spectra and bracketing B star observations following the methods of \cite{dalba20}. Recomputing the RVs using the best-match template that we identified increased the RV errors by a factor of $\sim2$, so we chose to continue to use the original CPS template. The poor match might be a consequence of Wolf 359's late spectral type---the library from \cite{dalba20} contains stars with $T_\mathrm{eff} > 3000$ K. Using the CPS template, RVs taken before the \keckhires detector upgrade in 2004 have a median measurement error of 8.2 m/s and post-upgrade RVs have a median measurement error of 3.9 m/s.

\subsubsection{MAROON-X}
We publish 68 measurements of Wolf 359 made with the MAROON-X spectrograph at Gemini Observatory.  The MAROON-X velocities are available in Appendix \ref{sec:RVappend} and online in a machine readable format.
The MAROON-X data were acquired with both the red (649--920\,nm) and blue (491--670\,nm) arm simultaneously during 34 observing nights. % for a total of 68 measurements. 
These observations were taken over 5 observing runs during February 2021, April 2021, May 2021, November 2021, and April 2022. 

Spectra were taken with a fixed exposure time of 30\,min and showed an average peak SNR of 90 pix$^{-1}$ in the blue arm and 460 pix$^{-1}$ in the red arm. The data were reduced by the instrument team using a custom \texttt{python3} data reduction pipeline to produce optimum extracted and wavelength calibrated 1D spectra. The radial velocity analysis was performed using SERVAL \citep{Zechmeister2018}, a template matching RV retrieval code in a custom \texttt{python3} implementation. On average, the RV uncertainty per datum was 1.0 m/s for the blue arm and 0.3 m/s for the red arm.
MAROON-X uses a stabilized Fabry-Perot etalon for wavelength and drift calibration \citep{MX2022} and can deliver 30 cm/s on-sky RV precision over short timescales \citep{Trifonov2021} but suffers from inter-run RV offsets with additional per-epoch uncertainties ranging from 0.5--1.5 m/s, corresponding to increased uncertainties of 1.4 m/s for the blue arm and 0.9 m/s for the red arm for signals on timescales longer than one month. 

\section{Analysis} \label{sec:analysis}

\subsection{Stellar Age Estimation \label{subsec:age}}

We provide an updated analysis of the age of Wolf 359 in order to constrain the sensitivities of our high-contrast imaging survey. We correlate our age estimates to our HCI survey sensitivity using evolutionary cooling models in order to determine the maximum mass of an unseen companion in Section \ref{subsec:ruledoutbyHCI}.   

\textbf{Gyrochronology:} The relation between rotation period, age, and mass has been studied extensively for low-mass stars \citep[e.g.][]{doi:10.1086/151310, doi:10.1088/0004-637X/721/1/675, doi:10.1088/0004-637X/727/1/56, doi:10.3847/1538-4357/abbf58}. It has been shown that stars begin their life with a fast rotation period and spin down with time via magnetic braking. The particular shape of this relation and the time it takes a star to spin down depends on its mass. The gyrochronology relation for Sun-like stars is calibrated, so the rotation period can be used to estimate an age. However, this gyrochronology relationship for Sun-like stars does not hold for M dwarfs \citep[e.g.][]{ doi:10.3847/1538-3881/ab3c53}. While the relationship for low-mass stars has not been  calibrated, it has been shown that rotation correlates with relative maturity \citep[e.g.][]{doi:10.3847/1538-4357/ac0444, doi:10.3847/1538-4357/ac90be, Pass2022}.

We calculated Wolf 359's Rossby number to be $R_0=0.02$ using the convective turnover time computed from \cite{Wright2011}. We then compared our $R_0$ value to Figure 6 in \cite{Newton2017}. We find that Wolf 359 lies in the magnetically saturated portion of this plot. 
For Sun-like stars, being in the saturated regime means the star is young ($<100$\,Myr). However M dwarfs stay rotating fast longer, thus a fast rotation period does not always mean the star is as young \citep[e.g.][]{doi:10.1088/0004-637X/727/1/56, doi:10.3847/1538-4357/ac77f9}. Recently, \citet{doi:10.3847/1538-4357/ac77f9} estimated that fully convective M~dwarfs transition between the saturated to the unsaturated regime at around $2.4\pm0.3$\,Gyr, which provides an approximate upper limit to the age of Wolf 359 but is not  informative. Below we combine rotation period with kinematics to estimate a more constrained upper limit on the age of Wolf 359.

\begin{figure}[ht]
\plotone{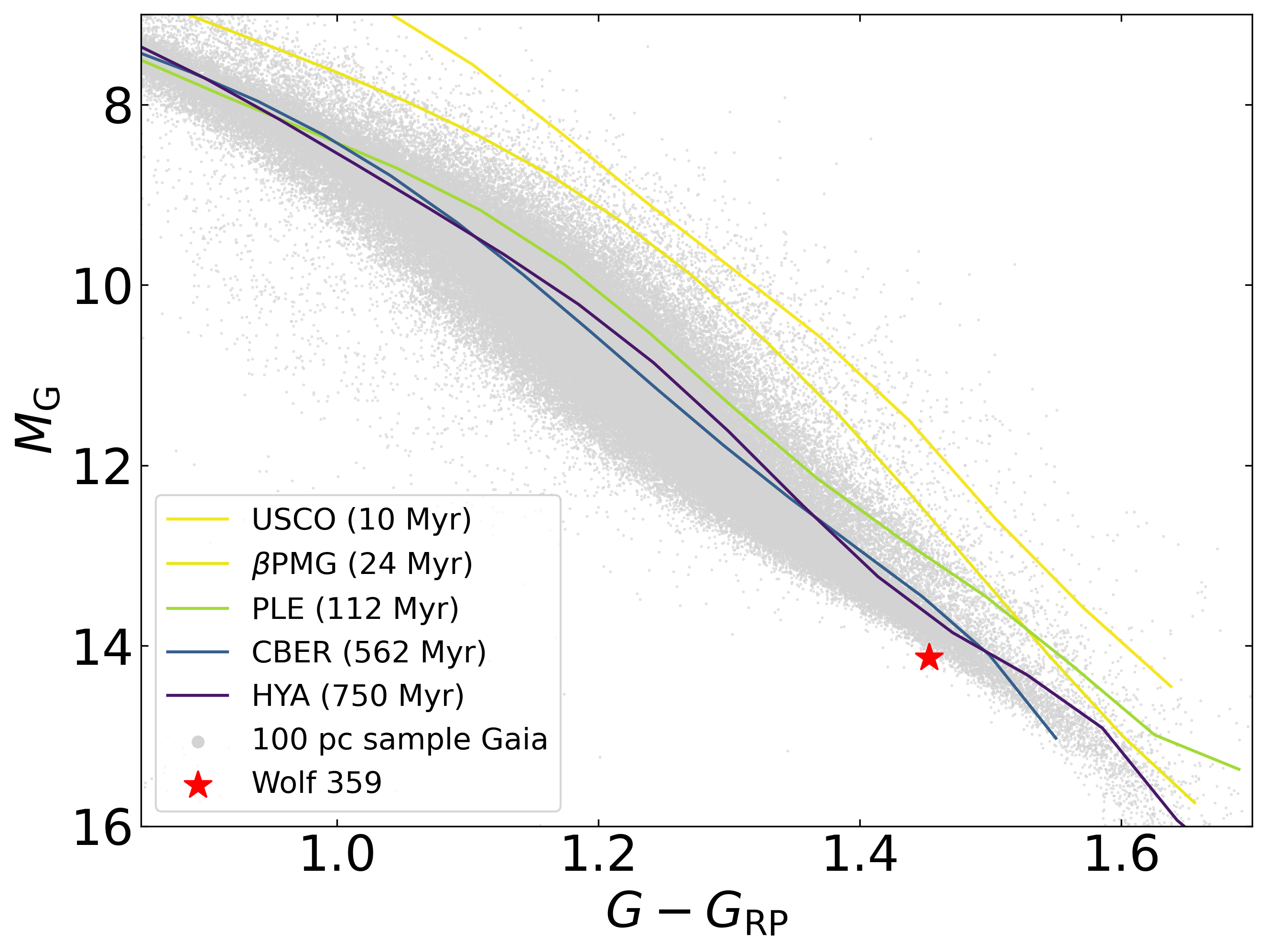}
\caption{CMD comparison with young moving groups:
We plot the color-magnitude diagram for Wolf 359 with empirical sequences from young associations of  ages 10\,Myr, 24\,Myr, 112\,Myr, 562\,Myr, and 750\,Myr \citep{doi:10.3847/2041-8213/ac0e9a} and the \gaia 100\,pc sample of M~dwarfs. The red star represents the position of Wolf 359. The color-magnitude position of Wolf 359 is not in agreement with the youngest moving groups of 10--112\,Myr. We find that Wolf 359 is in better agreement with the Coma Berenices (562\,Myr) and Hyades (750\,Myr) moving groups and the field sample. We conclude that Wolf 359 has converged on to the main sequence and that its age is older than $112$\,Myr.}
\label{fig:ageCMD}
\end{figure}

\textbf{CMD age dating}: We compared the color magnitude diagram position of Wolf 359 against the 100\,pc sample of M dwarfs from \gaia and empirical sequences based on bona fide members of young associations of several ages \citep{doi:10.3847/2041-8213/ac0e9a}. From Figure \ref{fig:ageCMD}, we conclude that Wolf 359 has already converged into the main sequence. This analysis suggests that Wolf 359 is older than the age of the Pleiades cluster (112\,Myr) as the lowest mass stars in this cluster have not converged into the main sequence.
From the CMD analysis, we conclude the age of Wolf 359 is older than 112\,Myr. 

\begin{figure}[ht!]
\plotone{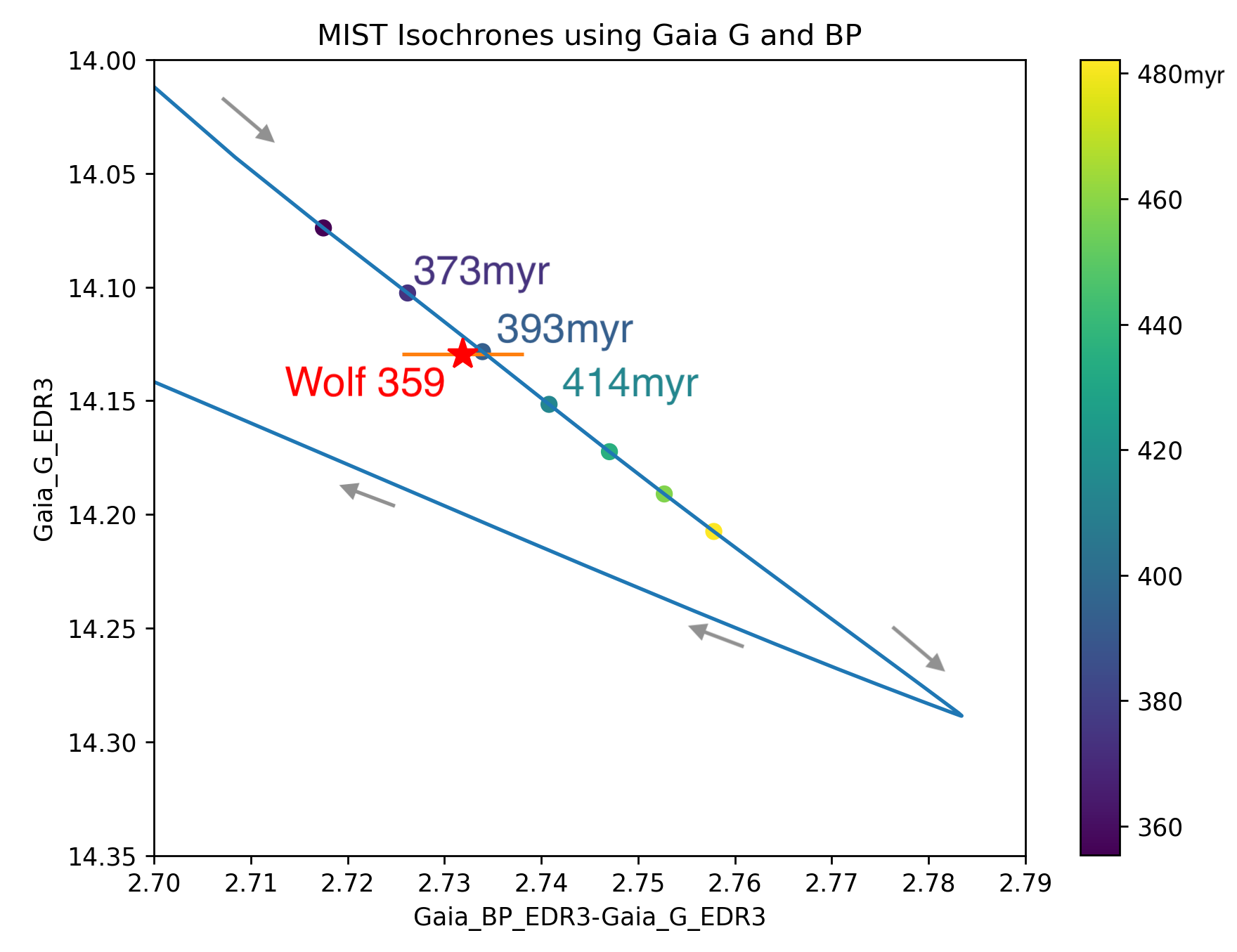}
\caption{Isochrone age dating: We used the MIST isochrone models with the \gaia eDR3 photometry in G and BP to estimate an age for Wolf 359. The blue line represents the MIST isochrone track for a star of $0.11$\,$M_\mathrm{\odot}$ with metallicity of $[Fe/H] = +0.25$\,dex.  Wolf 359 is represented by the red star, which lies closest to the isochrone point with an age of 393\,Myr (between 373\,Myr and 414\,Myr). We estimate an age of 400\,Myr from isochrone dating.}
\label{fig:MIST}
\end{figure}

\textbf{Isochrone age dating:} We used the MESA Isochrones and Stellar Tracks \citep[MIST;][]{MIST0, MIST} to estimate Wolf 359's age using a color-magnitude diagram. We adopt the MESA models associated with an M6 star ($0.11M_{\odot}$) with a metallicity of [Fe/H] $= 0.25$ dex \citep{Mann2015} and rotation of $0.4 v/v_\mathrm{crit}$. We used \gaia photometry (apparent magnitude $g = 11.038 \pm 0.003$, absolute magnitude $G = 14.130\pm 0.003$, apparent magnitude $g_{\rm bp} = 13.770 \pm 0.005$) to compare with the MIST isochrones (Figure \ref{fig:MIST}). 
Our iscohrone  age estimate is largely driven by the measurement of the \gaia G magnitude. 

While the MIST models can be unreliable for low-mass stars, they were recently shown to provide a good fit for stars like Wolf 359 with masses below $0.25\,M_{\odot}$ and a metallicity of [Fe/H]=+0.25 using the Hyades single star sequence \citep{Brandner2023}. We predict an age of $\sim400$\,Myr using the MIST models.

\begin{figure}[ht!]
\plotone{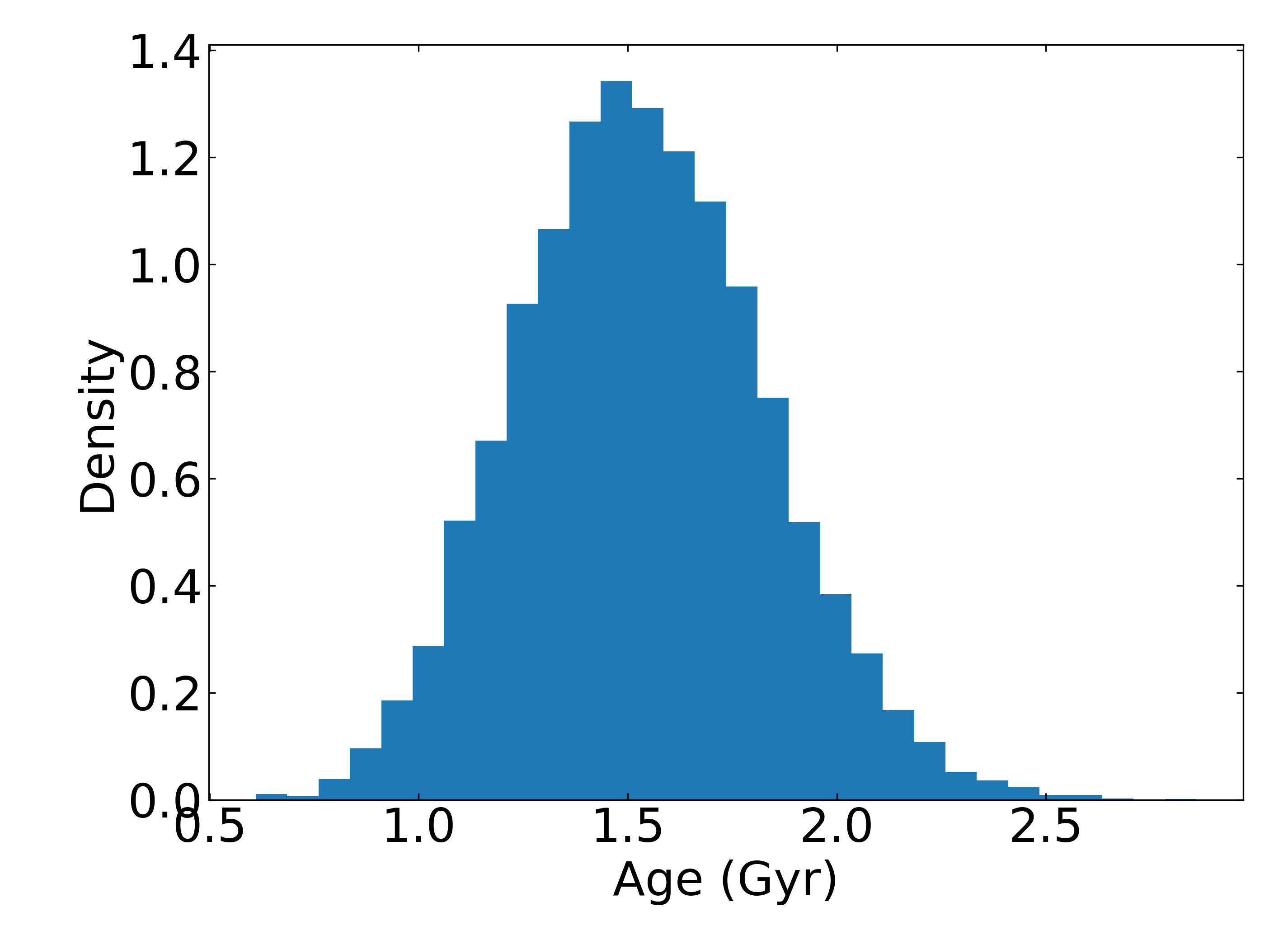}
\caption{Kinematic age dating: Wolf 359's kinematic age was measured using the methods outlined in \citealt{Kiman2019}. The results of the Monte Carlo simulation shown here finds the kinematic age to be $1.53 \pm 0.3$ Gyr. We adopt this kinematic age as our age upper-bound for Wolf 359.}
\label{fig:agekinematic}
\end{figure}

\textbf{Kinematic age dating:} We estimated Wolf 359's kinematic age to be $1.53 \pm 0.3$\,Gyr following the methods outlined in \citealt{Lu2021}. 
Briefly, this method consists of estimating the vertical velocity dispersion of a group of stars with similar temperatures and similar rotation periods. Assuming that the evolution of rotation period for stars with similar temperatures is the same, the stars in this group should have similar ages. Therefore we can use an age-velocity relation to estimate the average age of the group from the vertical velocity dispersion.
 We obtained a group of stars with similar mass and rotation period as Wolf 359 from the MEarth sample in \citealt{Newton2018}. 
We combined their reported rotation periods, mass, and radial velocities with their proper motions and parallaxes retrieved from \gaia eDR3 \citep{GaiaeDR32021} in \texttt{galpy}\footnote{\texttt{Galpy}: \url{https://github.com/jobovy/galpy}} \citep{galpy} to calculate their vertical velocities.   
We then created a bin in mass and rotation period around Wolf 359, selecting similar stars with similar ages.
To define the size of the bin, we used a group of stars with similar mass and rotation period as one M dwarf in the MEarth sample which is co-moving with a white dwarf. We used \texttt{wdwarfdate} \citep{Kiman2022} to get the age of the white dwarf from its effective temperature and surface gravity (retrieved from \citealt{GentileFusillo2021}), and set the bin size so the kinematic age of the group reproduced that age.
We used the age-velocity relation from \citealt{Yu2018} to correlate the vertical velocity dispersion with ages and then performed a Monte Carlo propagation of the vertical velocity uncertainties to determine the uncertainty in the kinematic age of Wolf 359. The resulting distribution from the Monte Carlo simulation is shown in Figure \ref{fig:agekinematic}. We obtained a kinematic age of $1.5 \pm 0.3$\,Gyr. However, as most of the stars in the bin are in the saturated regime, their rotation period still depends on their initial rotation period, making the dispersion in age larger. Therefore, we adopt an age of 1.5\,Gyr as an upper bound for Wolf 359’s age.

\textbf{Age summary:}  
Our age estimate from the MIST isochrone comparison ($400$\,Myr) is consistent with our young association comparison ($>112$\,Myr). Our CMD comparison with young moving groups shows it is probable that Wolf 359 has converged on to the main sequence.  While the 2.7\,d rotation period cannot be used to provide an exact age using gyrochronology, Wolf 359's fast rotation is a relative indicator of youth ($<2.4$\,Gyr).  We provide a better constrained upper bound estimate using the kinematic age dating of $\sim1.5\pm0.3$\,Gyr.   

For completeness through the remainder of this paper, we consider ages for Wolf 359 between 100\,Myr - 1.5\,Gyr in our HCI analysis. However, our analysis suggests that the ages estimated by 
\citealt{Pavlenko2006} using the spectral energy density distribution ($\sim 100-350$\,Myr) seem less likely due to Wolf 359's suspected convergence with the main group. 
If we someday measure the dynamical mass and temperature of an exoplanet companion around Wolf 359 using infrared direct imaging, we may then be able to apply planetary-mass isochrones to refine this age estimate.

\subsection{High-Contrast Imaging Analysis \label{subsec:ruledoutbyHCI}}

We used the Keck-NIRC2 contrast curves (Figure \ref{fig:cc}) to determine the final $5\sigma$ sensitivity of our imaging  survey across separations between 0.23\,AU and 4.18\,AU. 
We cannot make constraints on companions orbiting beyond separations of 4.18 AU on the night of observation 
because the field of view of the camera was limited to 5.1\arcsec $\times$ 5.1\arcsec (512$\times$512 pixel) to increase the speed of camera readout.

\begin{figure*}
\gridline{\fig{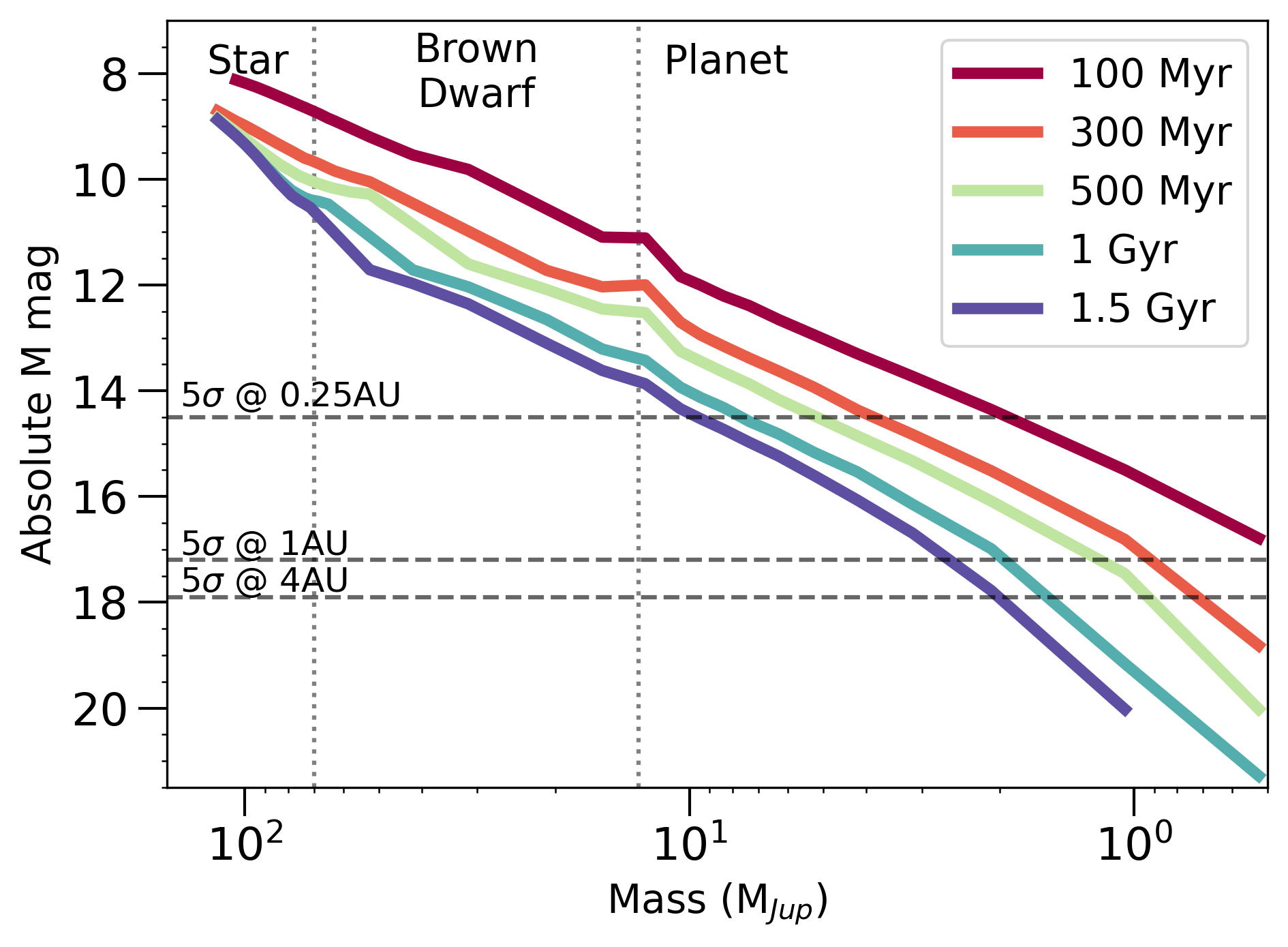}{0.45\textwidth}{(a) Baraffe Isochrones  for stellar, brown dwarf, and planetary cooling regimes}
\fig{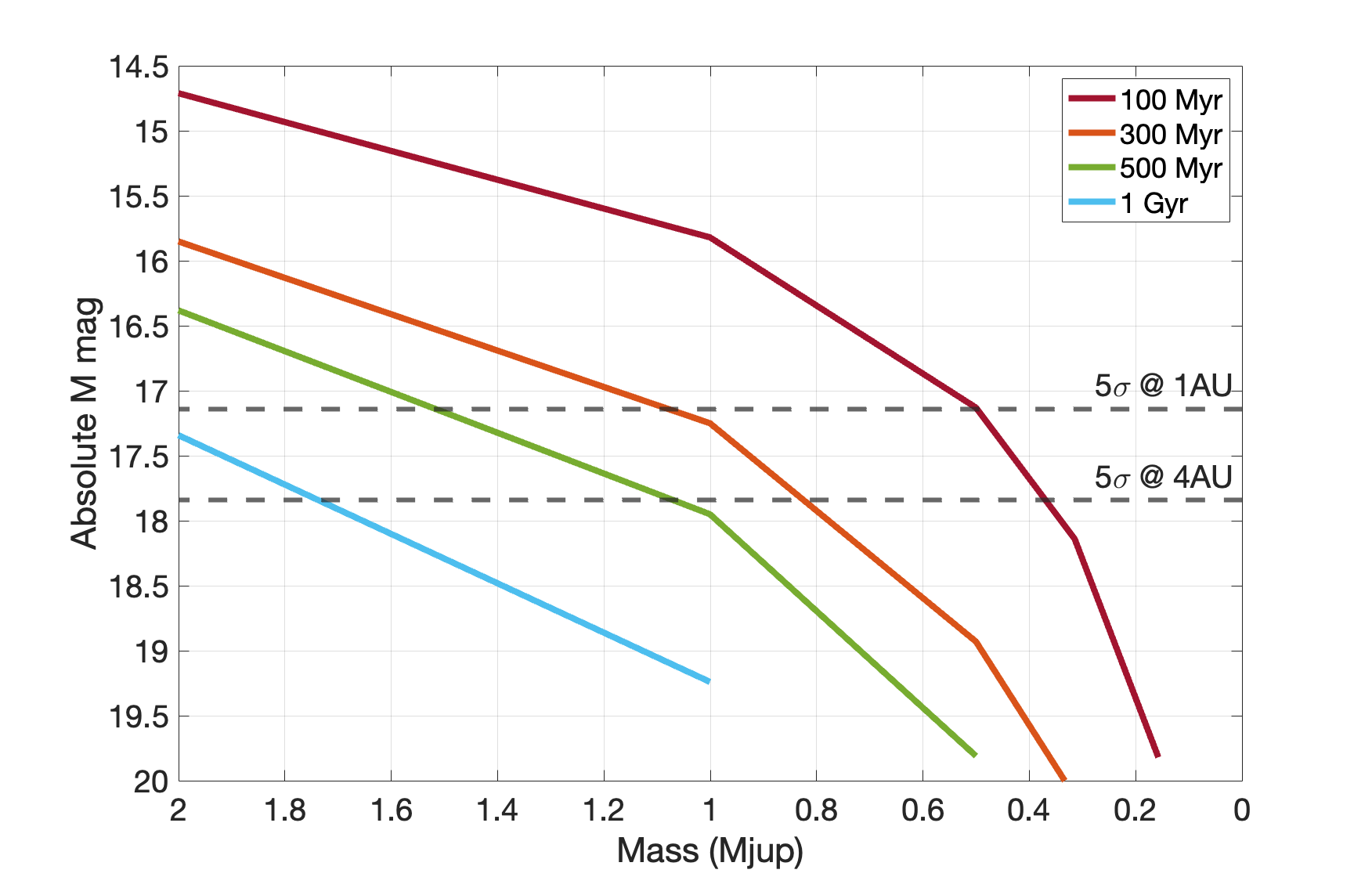}{0.5\textwidth}{(b) Linder+2019 low-mass planetary cooling curves}}
\caption{Isochrones overlaid with the $5\sigma$ constraints from the Keck-NIRC2 survey: The horizontal lines represent our imaging survey's $5\sigma$ sensitivity at 1 and 4\,AU of separation. (a) We use the BHAC15/DUSTY/COND models (\citealt{Baraffe15}, \citealt{DUSTY}, \citealt{COND}) to rule out all tight stellar and brown dwarf companions ($>13$\,\mjup) outside of 0.25\,AU (0.1\arcsec). (b) From 1--4.18 AU of separation, we apply the \cite{Linder2019} low-mass planetary cooling models to place upper mass limits on planetary companions. We rule out planets with masses $>1.5$\mjup to $5\sigma$ if Wolf 359 is younger than 500\,Myr in this region.     
\label{fig:COND}}
\end{figure*}

We then applied published isochrone models to predicted the upper mass limits for companions ruled out by the HCI observations.  
In Figure \ref{fig:COND}a, we applied the isochrone models created by Isabelle Baraffe\footnote{The Baraffe isochron models were retrieved at \url{http://perso.ens-lyon.fr/isabelle.baraffe/}.} to place constraints in the speckle-limited region at the tightest angular separations ($<1$\,AU). 
We used the BHAC15 models for the stellar regime \citep[\teff $> 3000$ K;][]{Baraffe15}, the DUSTY models for the brown dwarf regime \citep[1700 K $<$ \teff $<$ 3000 K;][]{DUSTY}, and the COND models for the planetary regime \citep[\teff $<1400$ K;][]{COND}. 
The Baraffe models predict that companions with masses above the deuterium burning limit ($>13$ \mjup) with ages younger $<1.5$ Gyr will be brighter than $M_s = 14.0$.  
Our survey reached a greater than $5\sigma$ sensitivity to companions with $Ms$ = 14 at separations greater than 0.25\,AU. We therefore rule out any stellar and brown dwarf companions orbiting outside of 0.25\,AU to 4.18\,AU at the time of observation. 

In Figure \ref{fig:COND}b, we used the isochrone models presented by \citealt{Linder2019} to set the mass upper limit in the background-limited regime from 1--4.18 AU (Figure \ref{fig:COND}b), where the sensitivity is limited by the sky background rather than the stellar contrast. Our combined-night contrast curve averages a sensitivity of $Ms = 17.7$ in this region. 
This sensitivity rules out companions with a mass bigger than 2.1 \mjup (667 \mearth) for ages younger than 1.5\,Gyr. We cannot rule out companions to $5\sigma$ with masses smaller than 0.4 \mjup (127 \mearth) for any adopted age older than 100\,Myr.

\begin{figure*}
\gridline{
\fig{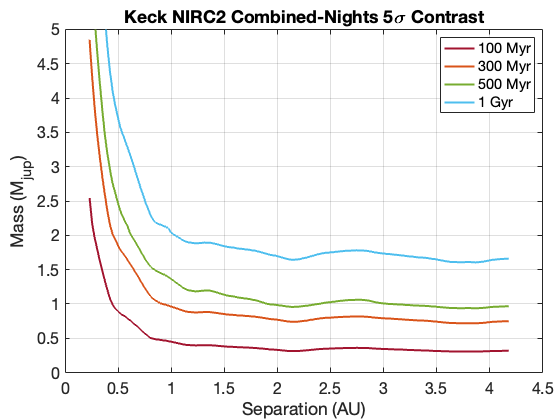}{0.45\textwidth}{(a) Keck-NIRC2 combined-nights $5\sigma$ contrast curve in mass-space for ages of 100\,Myr, 300\,Myr, 500\,Myr, and 1 Gyr}%{(a) Keck NIRC2 Combined Night }
\fig{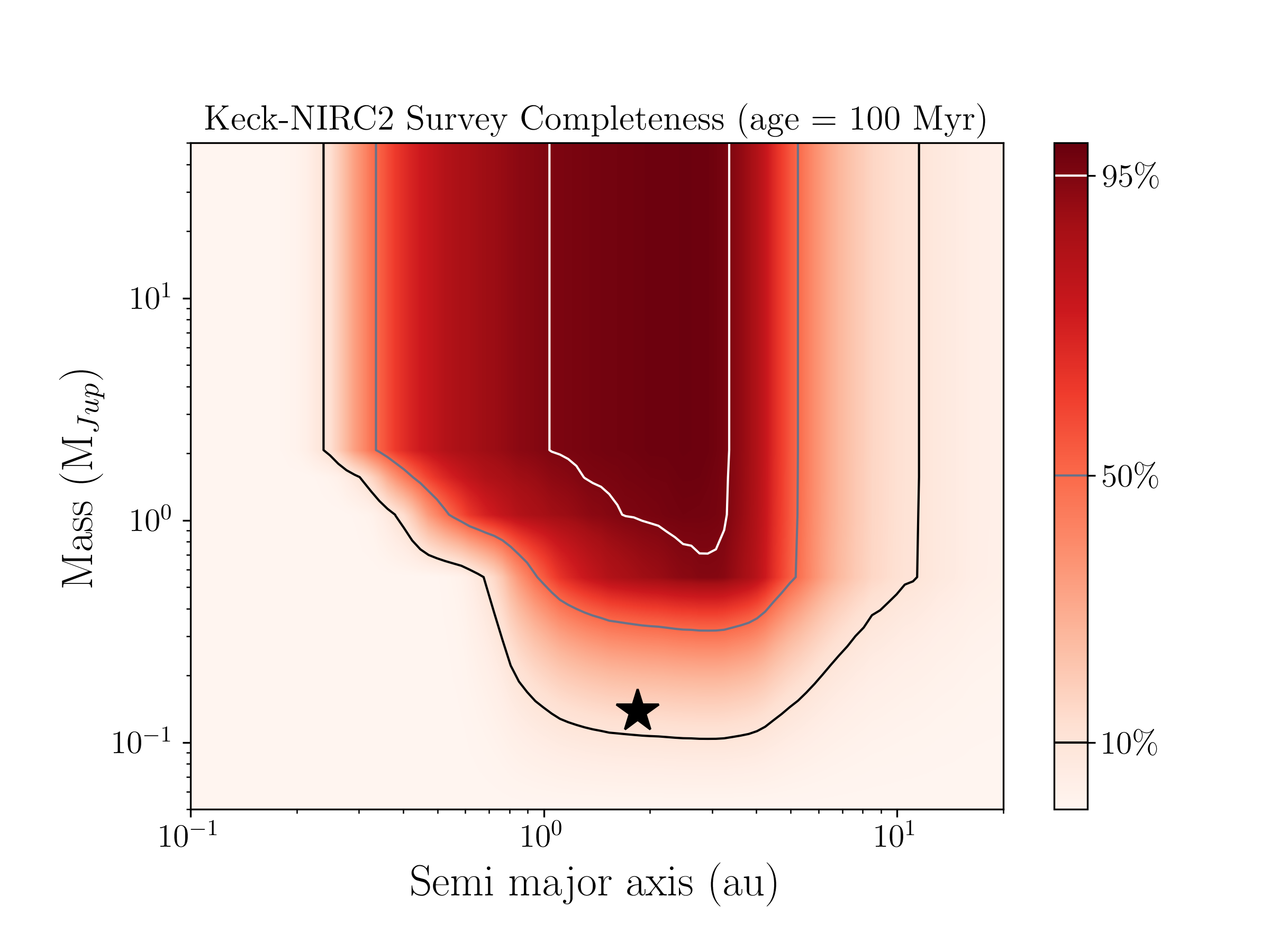}{0.5\textwidth}{(b) Completeness for an adopted age of 100 Myr} } %{(b) Linder+2019 low-mass planetary cooling curves}}
\gridline{
\fig{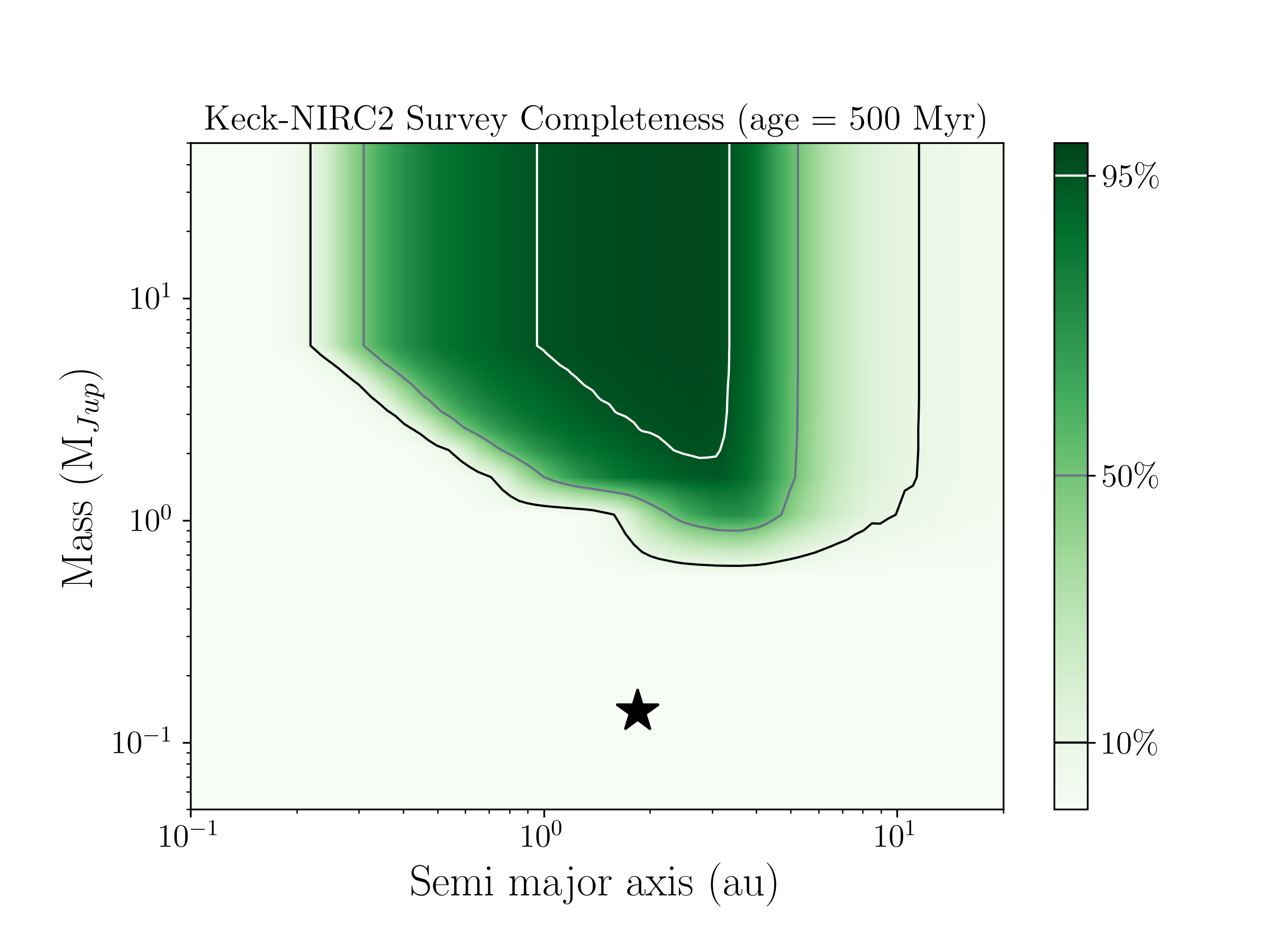}{0.5\textwidth}{(c) Completeness for an adopted age of 500 Myr} %{(b) Linder+2019 low-mass planetary cooling curves}}
\fig{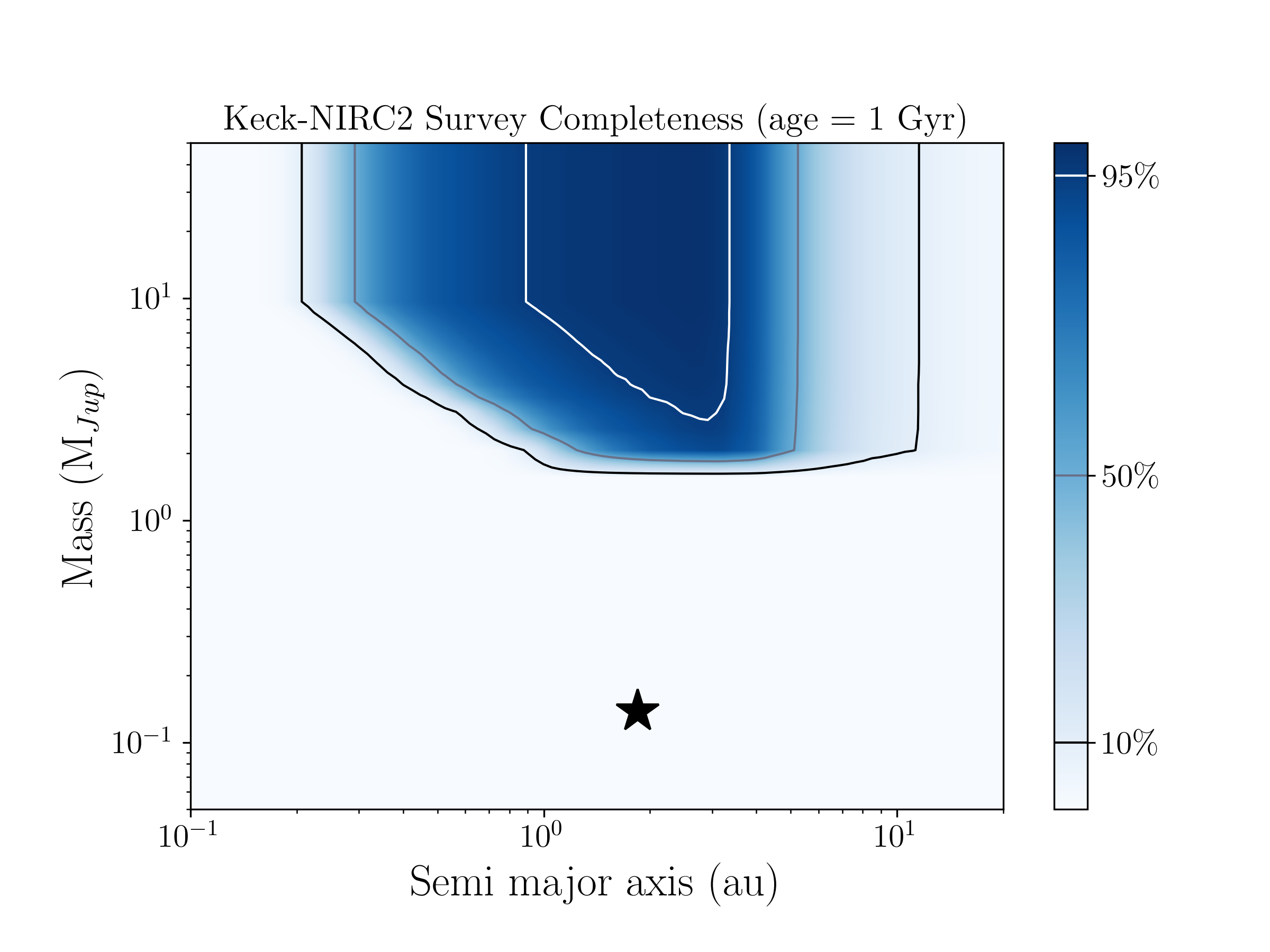}{0.5\textwidth} {(d) Completeness for an adopted age of 1 Gyr}}  %{(b) Linder+2019 low-mass planetary cooling curves}}
\caption{Keck-NIRC2 High Contrast Imaging Survey Completeness: (a) The NIRC2 combined-nights $5\sigma$ contrast was converted to mass space using the Linder+2019 and Ames-COND isochrone models.  (b-d) The NIRC2 survey completeness maps for the ages 100\,Myr, 500\,Myr, 1\,Gyr were estimated using the Exoplanet Detection Map Calculator (\texttt{Exo-DMC}) python package from the mass-space combined-nights $5\sigma$ contrast curves.   Our imaging survey has 10\% coverage to companions with a semi-major axis of 0.2-10\,AU and reaches 95\% coverage for companions with a semi-major axis between 1-3\,AU. The black star represents the Wolf 359 b semi-major axis and minimum mass as predicted by \citealt{Tuomi2019}.
\label{fig:exodmc}}
\end{figure*}

In order to estimate the completeness by mass and orbital semi-major axis of the high-contrast imaging survey, we utilized the Exoplanet Detection Map Calculator (\texttt{Exo-DMC}) package \citep{Bonavita2020} (Figure \ref{fig:exodmc}). 
We converted the combined-night $5\sigma$ Keck-NIRC2 contrast curves from apparent M mag into upper mass estimates adopting four ages: 100\,Myr, 300\,Myr, 500\,Myr, 1\,Gyr (Figure \ref{fig:exodmc}a). We used the Linder and Ames-COND isochrone models for this conversion and averaged the estimated masses in areas where the models overlapped. The Ames-COND isochrones\footnote{The Ames-COND models can be found at \url{https://phoenix.ens-lyon.fr/Grids/AMES-Cond/}} were accessed using the \texttt{species} package \citep{Stolker2020species}.

The greater than 10\% survey coverage spans from a semi-major axis range of 0.2\,AU to 10\,AU. We find the best survey coverage ($>95\%$) of semi-major axis between 1-3\,AU. Assuming an age younger than 1\,Gyr, we rule out companions with a semi-major axis of 1-3\,AU above $10\,M_{jup}$. While the semi-major axis predicted for the Wolf 359\,b candidate ($a=1.8 \pm 0.2$\,AU) is within this range, we do not reach the sensitivity to probe to the minimum mass predicted ($m\sin{i}\sim0.14\,M_{jup}$) regardless of age. For an age of 1\,Gyr, we rule out that the Wolf 359\,b candidate as described by \citealt{Tuomi2019} cannot be bigger than $4M_{jup}$. For an age of 100\,Myr, we rule out that the Wolf 359\,b cannot be bigger than $1M_{jup}$.

\subsection{Radial Velocity Analysis}
Our RV analysis incorporates 275 velocities from four instruments: Calar Alto Observatory's CARMENES \citep{CARMENES}, ESO-HARPS \citep{Mayor2003}, \keckhires, and Gemini-MAROON-X. 
The RV instruments and measurements used in our analysis of Wolf 359 are summarized in Table \ref{tab:RVsum} and are available in full in machine readable format online. The CARMENES data were retrieved from the DR1 release which spans from 2016--2020 \citep{Ribas2023}. The MAROON-X, HIRES, and HARPS data were provided directly by the observing teams. 

We elected to use the HARPS data as analyzed with the TERRA pipeline \citep{Anglada-Escud2012} in order to remain consistent with the analysis presented in \cite{Tuomi2019}. The 77 HARPS-TERRA velocities used in this analysis  incorporate the velocities presented in the 2019 announcement.  

The MAROON-X RVs were computed using both the red and blue arms of the spectrograph, producing two RV measurements per observation. We treat each the MAROON-X red-arm and blue-arm measurements as being from different instruments to account for different instrumental offsets and RV jitter amplitudes. We do the same for the \keckhires velocities collected before and after a detector upgrade in 2004. Within each instrument, we bin observations collected within 0.1\,d of one another. 
 
\begin{figure*}
\plotone{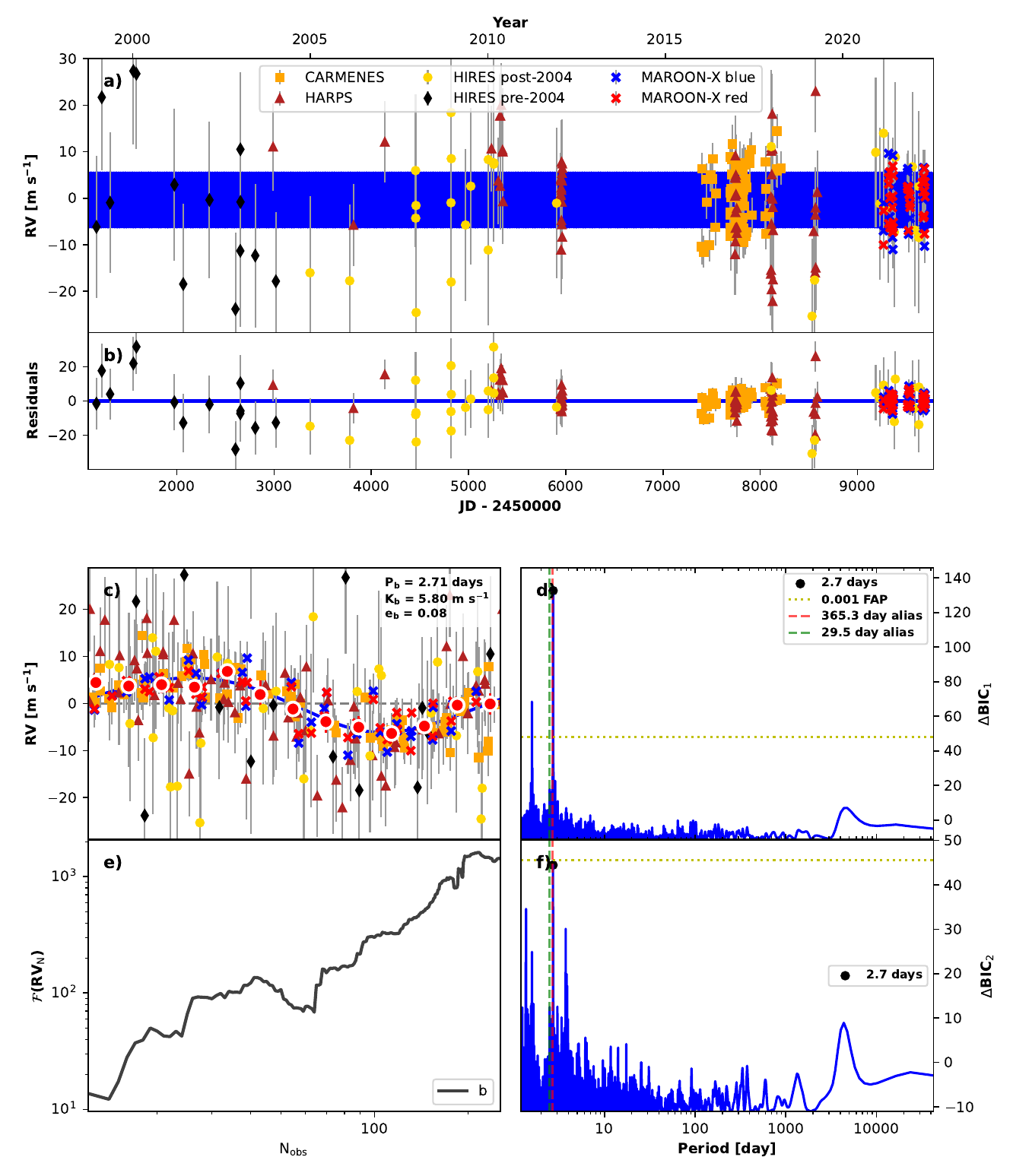}
\caption{\textbf{RV timeseries \& periodograms from analysis using RVSearch} \textit{(a)} We plot the RV timeseries using the available data from CARMENES, HARPS (TERRA pipeline), HIRES (pre- and post-2004), and MAROON-X (red and blue arm). The  blue line represents the detected signal from the known rotation period (2.71\,d; \citealt{Guinan2018wolf359rot}). \textit{(b)} The time series residuals after the stellar-rotation activity signal is removed. \textit{(c)} 
The folded timeseries for the rotation period signal. \textit{(d)} The periodogram before removing the rotation period signal. The highest peak corresponds to the rotation period signal (2.71\,d), and the second peak corresponds with half the rotation period (1.4\,d). \textit{(e)} We find that the quantification of the strength of the detection for the 2.7 d signal as a function of the number of observations monotonically increases as expected.  
\textit{(f)} The periodogram of the residuals after removing the 2.71 d signal from the stellar rotation period. We do not find evidence for any additional candidates above our False Alarm Probability threshold ($0.1\%$). The local peak at 4370\,d corresponds to a $\Delta BIC = 7.0$ (0.001 FAP corresponds with $\Delta BIC=45.7$).  
\label{fig:RV}}
\end{figure*}

\begin{figure*}
\gridline{\fig{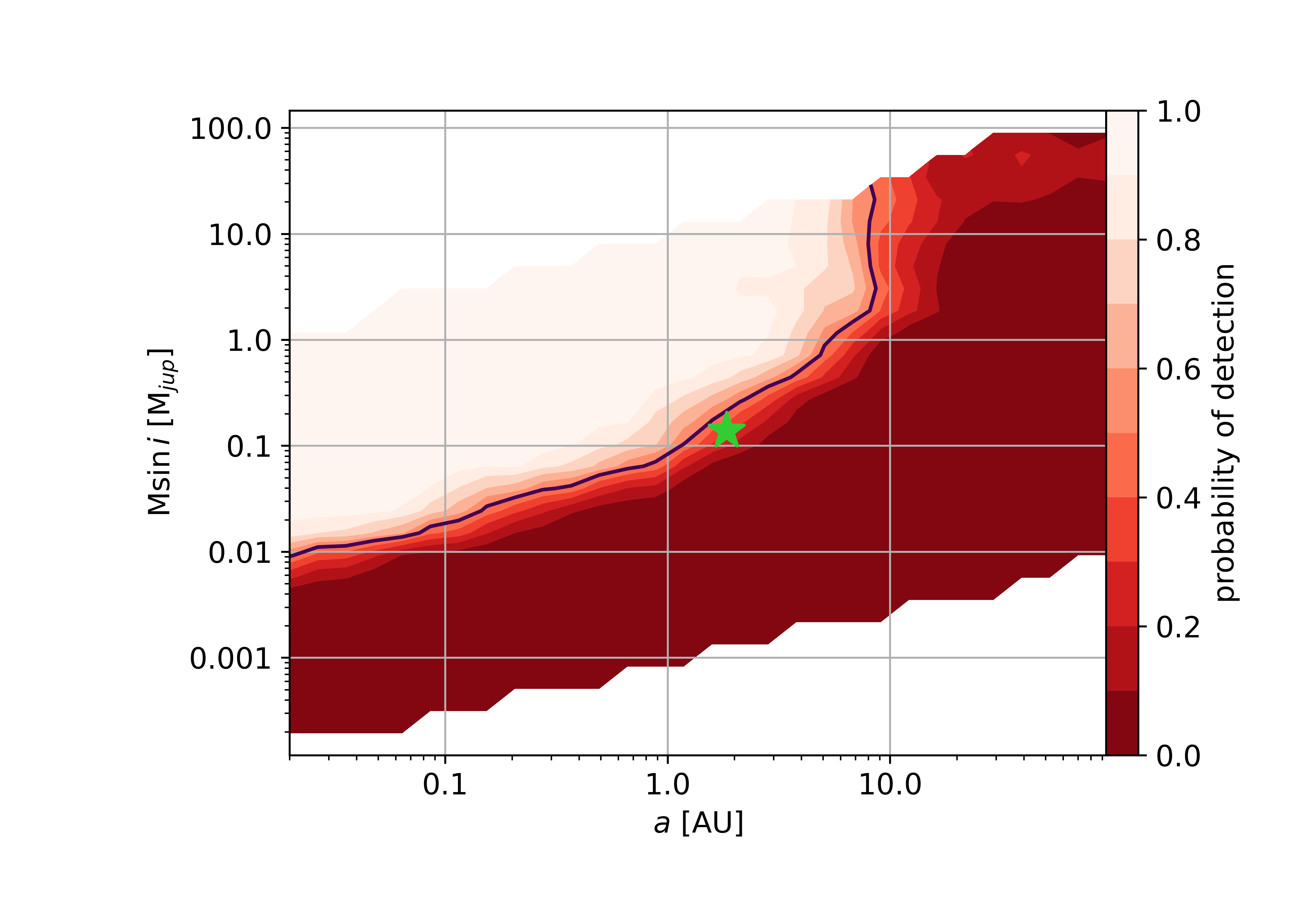}{0.52\textwidth}{(a) Wolf 359 Radial Velocity Survey Completeness Contour}
\fig{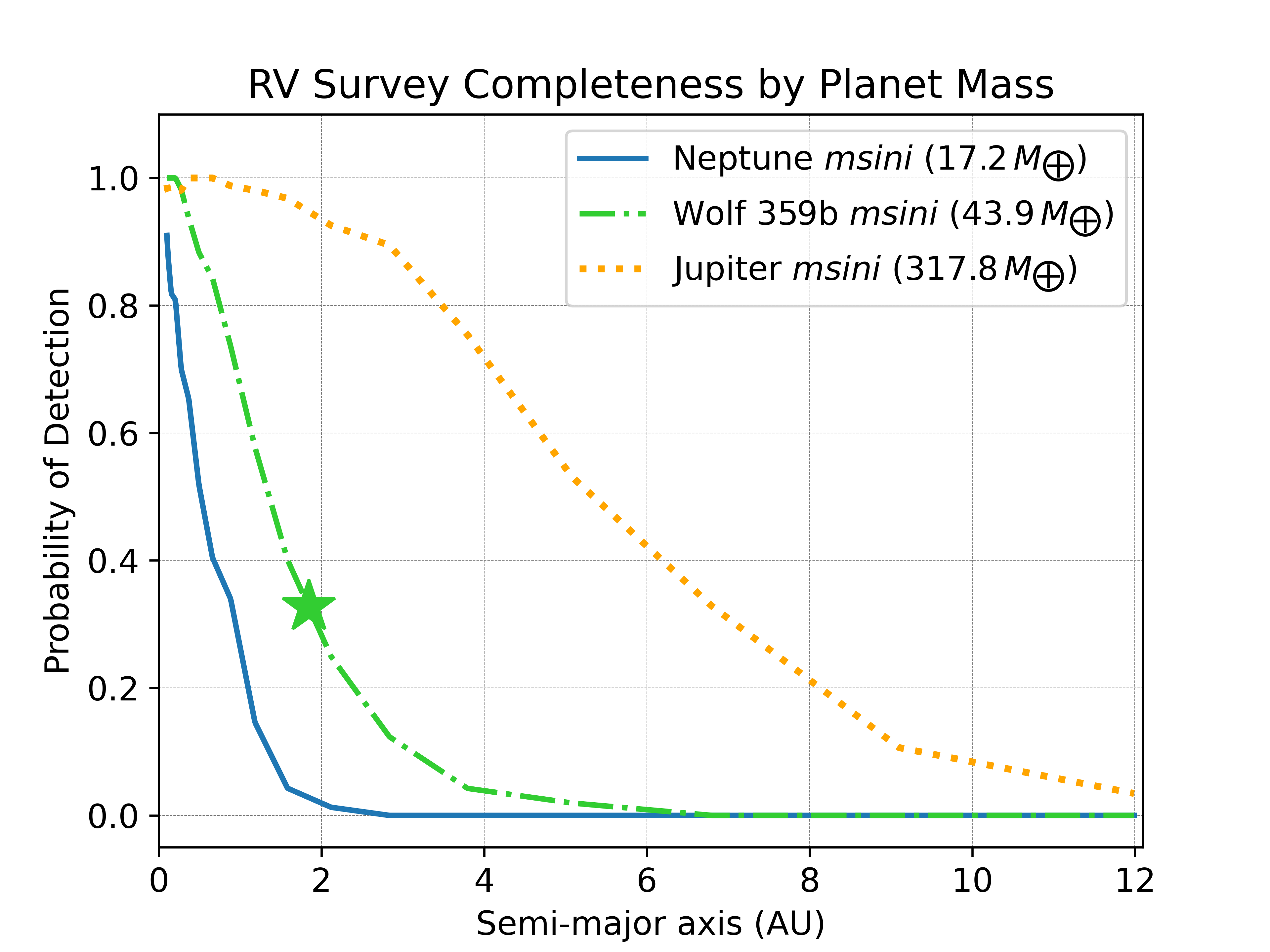}{0.47\textwidth}{(b) Probability of Detection by Minimum Mass}}
\caption{Radial Velocity Survey Completeness:  We used the injection-recovery function within \texttt{RVSearch} to determine the completeness of our Wolf 359 RV survey as a function of the minimum planet mass and semi-major axis.  Our analysis methods yield a 32\% chance of recovering a signal that matched the Wolf 359b candidate as described by \citealt{Tuomi2019} (green star). 
\label{fig:RVcompleteness}}
\end{figure*}

We used the \texttt{RVSearch}\footnote{\texttt{RVsearch}: \url{https://github.com/California-Planet-Search/rvsearch}} python package \citep{rosenthal21} to perform a blind planet search within our RV timeseries data (Figure \ref{fig:RV}).  
We detected the known signal associated with the rotation period of the star (2.71 d). Once the stellar-rotation activity signal was removed, we detected no signals over a False Alarm Probability of $0.1\%$. 
We used the injection-recovery tools built into \texttt{RVSearch} to estimate the sensitivity of our RV survey to planets of specified $m\sin{i}$ and semi-major axis to create the completeness contour shown in Figure \ref{fig:RVcompleteness}. 
The probability of detection for a planet with a minimum mass equivalent to a Neptune-mass, Jupiter-mass, and the Wolf 359\,b candidate is also shown in Figure \ref{fig:RVcompleteness}.   
\texttt{RVSearch} yielded a 32\% completeness to an equivalent $m\sin{i}$ and semi-major axis as the Wolf 359\,b candidate. Because we do not have a significant completeness in this space, we are not able to confirm or deny the candidacy of Wolf 359\,b using \texttt{RVSearch} with our RV dataset.   

To further explore the candidacy of Wolf 359\,b, we used the open-source software package \radvel\footnote{\texttt{Radvel}: \url{https://github.com/California-Planet-Search/radvel}} \citep{radvel} to model the RV data. We used the \citealt{Tuomi2019} results for Wolf 359b listed in Table \ref{Tab:ExoplanetCandidates} as priors.  We employed fits with and without the Gaussian Process Fitting module which can be used to fit and remove signals due to stellar activity. We ran our \radvel MCMCs using $N_{walkers} = 50$, $N_{steps} = 10000.0$, $N_{ensembles} = 6$, and $Min Auto Factor = 30.0$.  In all \radvel fits, the chains did not pass the convergence test to indicate that the walkers were well mixed. The convergence criteria could not be met, so we draw no conclusions about the properties of the Wolf 359b candidate from our \radvel fits. 

We detected no new candidates. 
At 95\% confidence, our RV analysis excludes planets with a minimum mass bigger than \mpsini $> 13.5$ \mearth (0.04 \mjup) for $a=0.1$\,AU and planets with a minimum mass bigger than \mpsini $> 147$ \mearth (0.46 \mjup) for $a=1$\,AU. We have over 50\% completeness to exclude planets with an \mpsini equivalent or bigger than 1\mjup within 5.3\,AU and 1 Neptune-mass within 0.52 AU.  
Our RV survey has little coverage to companions orbiting with a semi-major axis larger than $a > 10$ AU for all masses.

\begin{deluxetable*}{c c c c c c c}
\tablecaption{Wolf 359 RV data summary. \label{tab:RVsum}}
%\tabletypesize{\largesize}
\tablehead{\colhead{Instrument} & \colhead{Source} &\colhead{\begin{tabular}[c]{@{}c@{}}Spectral \\ Range (nm)\end{tabular}} &
\colhead{\#Meas.} & \colhead{Baseline} & 
%\colhead{STD of RV} & 
\colhead{\begin{tabular}[c]{@{}c@{}}Avg. RV \\ Precision\end{tabular}} & \colhead{\begin{tabular}[c]{@{}c@{}} Inst. \\ Offset*\end{tabular} } }
\startdata
CARMENES & Retrieved from \cite{Ribas2023} & 550-1700 & 78 & 2.23 yr   & 1.99 m/s  & 0.05 m/s   \\
ESO-HARPS      & Directly from M. Tuomi & 378-691 & 77     & 15.3 yr  & 3.09 m/s   & -3.22 m/s \\
HIRES Pre-2004  & California Planet Search team   & 300-1000  & 14 & 5.05 yr  & 8.09 m/s & -9.68 m/s                                                  \\
HIRES Post-2004 & California Planet Search team & 300-1000  & 38  & 17.14 yr & 4.26 m/s  & -3.28 m/s   \\
MAROON-X blue   & MAROON-X team   & 499-663 & $34^{\dagger}$ & 1.17 yr  & 1.39 m/s  & -6.47 m/s \\
MAROON-X red    & MAROON-X team  & 649-920 & $34^{\dagger}$ & 1.17 yr  & 0.88 m/s  & -5.62 m/s  \\  
\enddata
\begin{tablenotes}
      \small
      \item $^\dagger$The MAROON-X blue and red data were collected simultaneously
      \item * The instrument offsets were calculated from the fit made using \texttt{RVSearch} when detecting the signal from the stellar rotation period.
\end{tablenotes}
\end{deluxetable*}

\section{Discussion} \label{sec:discussion}

\subsection{Performance of the direct imaging survey with Keck-NIRC2}

Few Keck-NIRC2 HCI observations have been published that span multiple nights that utilize the Ms filter ($4.67\mu$m) in conjunction with the vector vortex coronagraph.   
Previous published deep surveys of this type have so far limited to 
the Eps Eri results from \citealt{Mawet2019} and \citealt{Llop-Sayson2021}.  
However, it is expected that surveys similar to the work presented here 
will become more common as data from indirect methods of exoplanet detection become more widely available and drive targeted direct imaging surveys towards studying colder companions.  
We document the expected performance of our imaging survey as compared to our measured performance to aid in the planning of future multi-night Keck-NIRC2 HCI surveys that are completed with the Ms filter with the vortex coronagraph.

We report that our measured efficiency on the night with the greatest number of images (2021 March 31 UT) was 52\%. This excludes the setup time and used the observing configuration described in Section \ref{subsec:obsHCI}. After our initial setup, we observed Wolf 359 for 4.05\,hr and totalled 2.12\,hr of science integration time. We ran the majority of QACITS sequences with 50 science images (22.5\,min total integration time) and experienced no significant QACITS centering issues while collecting science data.

We adopt the predictions produced by the Keck Observatory's online 
\href{https://www2.keck.hawaii.edu/inst/nirc2/nirc2_snr_eff.html}{NIRC2 SNR and Efficiency Calculator} to quantify the expected SNR in the background limited regime of our contrast curves.  These equations for the NIRC2 SNR Calculator are outlined in Appendix \ref{sec:revisedcalc}. 
We do not consider the speckle-limited regime of our contrast curves from this comparison ($sep < 0.8\arcsec$) as the NIRC2 SNR calculator cannot quantify the SNR in the speckle-limited region. 

We evaluate the performance using one night of observations to avoid complications in the performance discussion from combining data across multiple nights. We elected to use 2021 March 31  (UT) because it is the night of our survey with the most available data. 
 Our contrast curve for this night was generated using 269 of the 283 images taken with an exposure time of 0.3\,s and coadd of 90, totalling approximately 2\,hr of integration time. We measured an average $5\sigma$ contrast in the background-limited region of the contrast curve ($>0.8$\arcsec) to be $\Delta m_s$ = 8.53 (apparent magnitude of $m_s =14.38$). 
 
 Our measured $5\sigma$ detection limit from 2021 March 31 is consistent with the performance on individual nights of the Eps Eri survey 
where \citealt{Llop-Sayson2021} used the pyramid wavefront sensor (pyWFS) to collect approximately 2 hours of integration time.
The best SNR achieved by \citealt{Llop-Sayson2021} was between the separations of $1.5\arcsec - 1.75\arcsec$ and corresponds to an apparent magnitude of $m_s = 14.4$ ($\Delta mag = 12.7$). Both this work and the Eps Eri surveys indicate it is improbable to detect a companions dimmer than $m_s = 14.4$ to $5\sigma$ with this instrument configuration in one half-night of Keck NIRC2 time when operating with the vortex coronagraph paired with the pyWFS.  

We next checked our measured results against the prediction made by the NIRC2 SNR calculator using the parameters that matched our observing setup: 0.3\,s integration time with 90 coadds, narrow mode, 2 reads, 269 images, and no telescope nodding. We assumed a Strehl ratio of 0.85 which is a conservative estimate associated with 300\,nm of wavefront error. The NIRC2 SNR calculator assumes that the background flux and flux from the source will follow Poisson statistics.  We find that the calculator predicts the $5\sigma$ threshold to be at an apparent magnitude of $m_s = 16.07$, which is not consistent with with our observed results. Our measured SNR was 1.69 magnitudes brighter than the predicted performance by the NIRC2 SNR calculator, meaning we were more restricted in the companions that we could detect at the background-limited wide separations than was predicted by the calculator. 

We expect that the prediction by the NIRC2 calculator would be somewhat inconsistent with our results because the NIRC2 SNR calculator was not designed to predict observations when the vortex mask is used. To better refine our predicted performance estimate, we modified the equations used by the NIRC2 SNR calculator.  These modifications are documented in Appendix \ref{sec:revisedcalc} and incorporate a throughput penalty to the measured signal to account for the use of the fixhex pupil stop and the vortex mask at 4.7\,$\mu m$ (Total throughput penalty, $0.57\pm 0.03$). 
We additionally offer a revision to the background flux counts when the vortex is used in M-band (17850 DN/s per pixel). 
 When we apply our revised equations to estimate the predicted performance for our 2021 March 31  dataset, we find that our $5\sigma$ detection threshold is predicted to be at $m_s=15.49$. %$m_s= 15.26$.
 While this estimate better aligns with our measured performance, this method still over-estimates the brightness of the $5\sigma$ detection threshold by 1.11 magnitudes when comparing to our measured performance from that night ($m_s=14.38$). 

We ruled out the possibility that this performance gap was due to uncorrected non-uniform background counts spatially in the individual images through highpass filtering. 
We used \texttt{VIP}'s internal highpass filtering function
to determine the optimal highpass filtering by injecting a fake planet into each individual image, running six types of highpass filters on each image, and then using stellar photometry to recover the SNR of the injected planet.  The optimal highpass filtering method was \texttt{gaussiansubt} with size $2.25*fwhm_{nirc2}$. 
We then edited \texttt{VIP}'s contrast curve function to include the highpass filtering step using the optimal highpass filter. The highpass filtering step was added after fake planet injection but before running PCA.  There were slight differences between the contrast curves produced from the image sets with and without the highpass filter, but the differences did not affect the contrast achievable in the background limited region of the image. We thus conclude that the performance gap is not due to poorly corrected background structures in each frame.  

To determine if the performance gap was due to the image background noise not obeying Poisson statistics temporally, we measured how the sky background noise over time compared to the statistics expected from photon noise. 
 We measured the sky background noise by summing the counts inside four circular apertures with a diameter equal to the $fwhm_{NIRC2}$ using the 2021 March 31 image set before and after sky subtraction was completed.  The apertures were located 1.76 arcseconds from the image center in the direction of the image corners in order to avoid contamination from the star.  We found our measured background noise value using 20 frames from the image cube after the sky subtraction was applied. The 20 frames were chosen from the full cube where the conditions were stable (no background drift, average background counts in the raw frames are consistent, and similar adaptive optics correction). We plotted the aperture sum counts of each aperture and then took the standard deviation of the counts over time.  The corresponding photon noise value was determined using the image cube before the background subtraction was made. We measured the sum of photons inside each aperture, averaged the sums, and then took the square root of the average sum to act as the expected photon noise. 
The ratio between our measured-noise to photon-noise contribution was 1.9 from the subset of the 20 stable frames.  Across the full image cube, we found the ratio of measured/theoretical-photon noise to be 3.0. This corresponds to a flux difference of 0.69 and 1.2 magnitudes respectively. This range of values is consistent with the performance gap we see after accounting for the throughput loss from the vortex and pupil stop ($\Delta mag = 1.1$).

We hypothesize that the background noise does not follow Poisson statistics because of short time-scale water vapor variations at 
timescales less than the length of our 30\,s images.  This hypothesis could be tested when upgrades to the NIRC2 electronics are completed in 2023 which will allow for faster readout and background corrections to be made at shorter timescales. If proven true, the limits of previous surveys may be improved upon by observing the target again using sub-second integration times in order to improve background correction.

\subsection{Prospects for Directly Imaging an Exoplanet around Wolf 359 using JWST} 

JWST offers an opportunity to directly image exoplanets in infrared wavelengths without the contamination from the Earth's atmosphere allowing for the telescope to probe for colder companions as compared to ground based telescopes. 
In this section, we present simulations to explore the potential of JWST to directly image a cold giant planet orbiting Wolf 359 using the Near Infrared Camera (NIRCam) Coronagraphic Imaging mode and Mid-Infrared Instrument (MIRI) imaging. MIRI and NIRCam can be used in combination to span wider coverage for companions in orbital separations, cloudiness, and temperature.  NIRCam can be used to achieve high contrasts at sub-arcsecond inner working angles at shorter infrared wavelenghts (0.6-5$\mu$m, \citealt{Rieke2023}), which was demonstrated successfully during the Early Release Science Program to image the super-Jupiter mass exoplanet HIP 65426b \citep{Carter2022}.  MIRI operates at longer infrared wavelengths (5-28$\mu$m, \citealt{Wright2023}), giving greater sensitively to cold and cloudy companions. 

Because of Wolf 359's proximity, a planet revealed through NIRCam or MIRI imaging has the potential to become the coldest directly image exoplanet that could be characterized with JWST spectroscopy.  If such an exoplanet is detected, detailed characterization would allow the planet to become an anchor to test theories related to the atmosphere and formation of cold gas giant and ice giant planets.

\subsubsection{NIRCam Coronagraphic Imaging}

We explore the possibilities of using the NIRCam Coronagraphic Imaging mode to directly image companions orbiting Wolf 359 by simulating contrast curves using the Pandeia Coronagraphy Advanced Kit for Extractions (\texttt{PanCAKE}) python package\footnote{Pandeia Coronagraphy Advanced Kit for Extractions; \url{ https://github.com/spacetelescope/pandeia-coronagraphy}} (\cite{Girard2018}, \citealt{Perrin2018}, \citealt{CarterPancake}). We considered observations in the F444W filter, as the broadest band between the 4-5$\mu$m peak in brightness, in conjunction with the round coronagraphic mask MASK335R. We simulated integration times of 20\,min, 1\,hr, and 10\,hr with ADI and RDI subtraction techniques. 
To simulate the ADI contrast curve, we assumed the total exposure time was split between two rolls (0$^{\circ}$ and 10$^{\circ}$) when imaging the target.  For the RDI simulations, we assumed a perfect reference with the same properties of Wolf 359 and used a 9-point circle dither pattern.  PSFs were generated using the precomputed library over on-the-fly generation with wavefront evolution to reduce computational intensity. As such, these contrast curves represent an optimistic estimate of the achievable performance. We allowed PanCAKE to optimize the readout parameters for dither pattern, number of groups, and number of integrations.

\begin{figure}[ht!]
\plotone{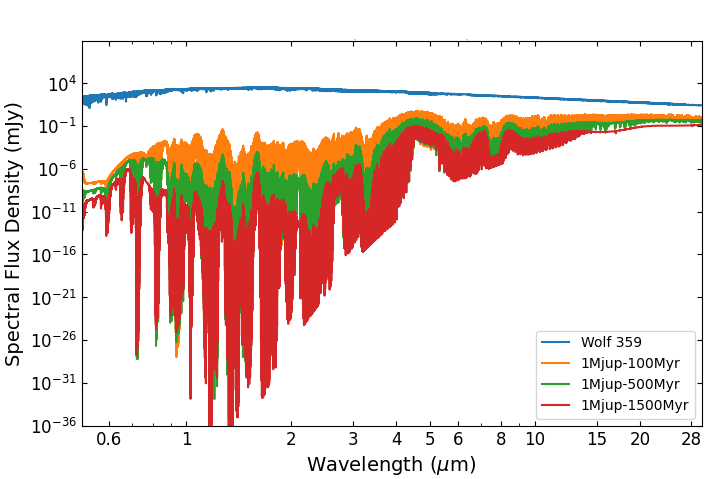}
\caption{Simulated atmospheric models for Wolf 359 and a cloudless 1\mjup companion: The modeled companion spectra shown correspond to ages of 100\,Myr (orange), 500\,Myr (green), and 1.5\,Gyr (red).  The simulated spectra of the Wolf 359 host star is shown in blue.  The estimates of the flux between $3.881\mu m-4.982\mu m$ were used to determine the expected brightness and expected SNR for each companion type to simulate a NIRCam observation with F444W + MASK335R using \texttt{PanCAKE}.   }
\label{fig:Sagnickmodels}
\end{figure}

To estimate what types of exoplanets may be detectable, we generated  atmospheric models for companions with masses between 20 \mearth - 1 \mjup for ages spanning 100\,Myr - 1.5\,Gyr using the \texttt{PICASO 3.0}\footnote{Planetary Intensity Code for Atmospheric Spectroscopy Observations; \url{https://github.com/natashabatalha/picaso}} \citep{Mukherjee23,Batalha2019} radiative--convective--thermochemical equilibrium model to simulate cloud-free 1D atmospheres for such companions. We assumed solar metallicity and C/O ratio for our simulated atmospheres. To estimate the $T_{\rm eff}$ and radius of a companion with a given mass at a certain age, we used the \cite{Linder2019} evolutionary tracks and linearly extrapolated along the age axis when needed.  The Phoenix stellar models \citep{Husser2013} were employed to generate the stellar model for Wolf 359 using a spectral type of M5V and the Vega mag scaled to $2MASS\,k_s = 6.084$. 
An example of the set of thermal emission spectra from our generated atmospheric models are shown in Figure \ref{fig:Sagnickmodels}.

\begin{figure*}[ht!]
\plottwo{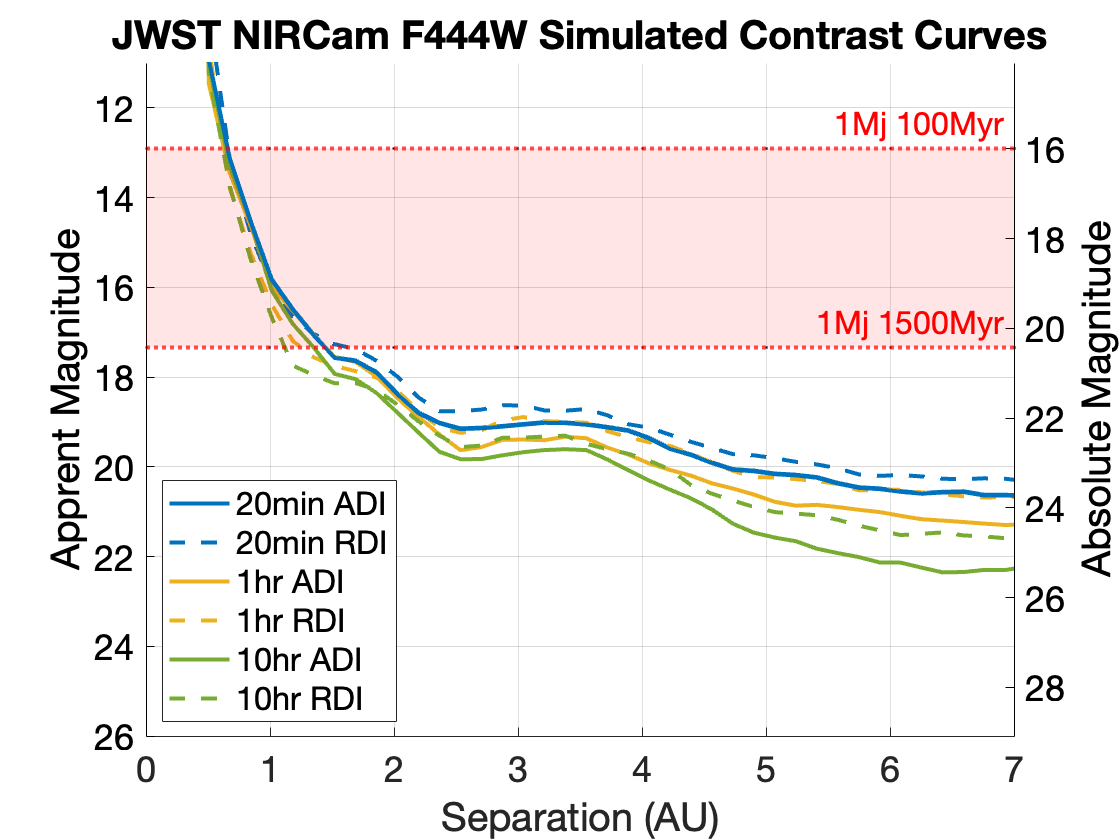}{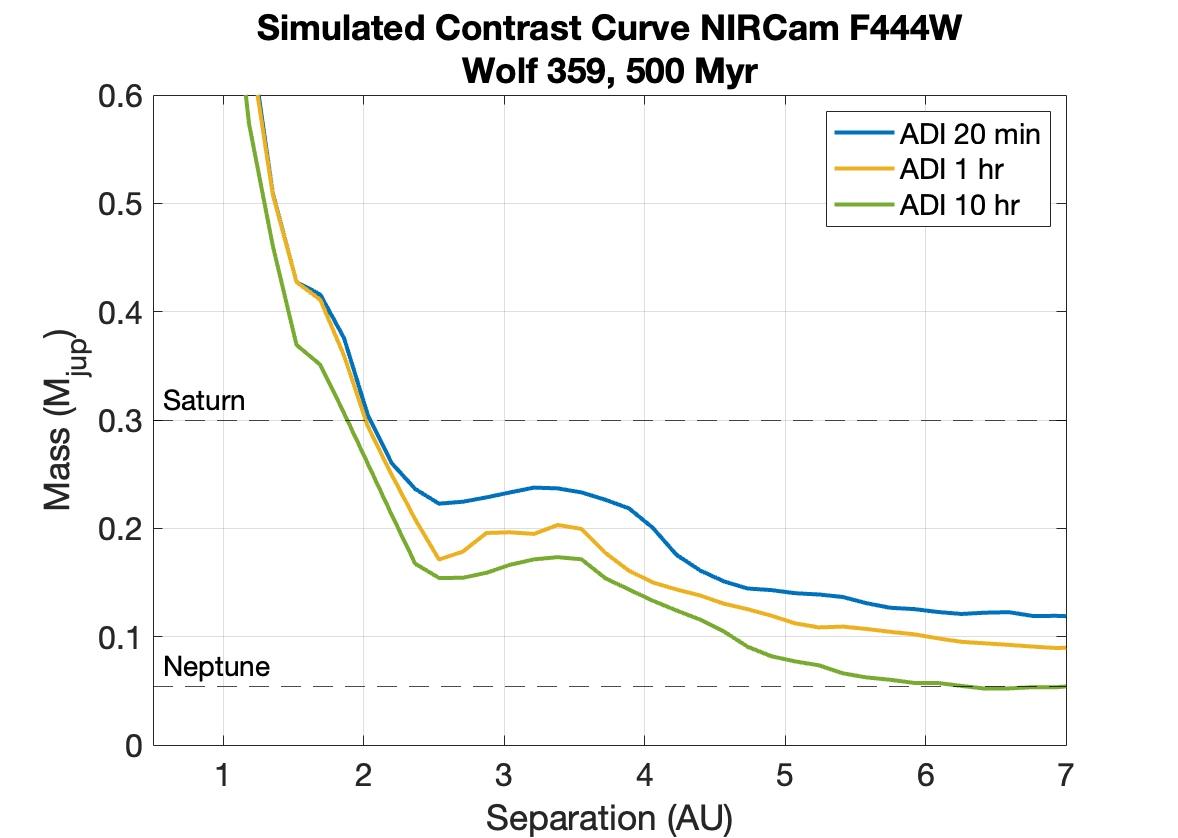}
\caption{Simulated JWST NIRCam Coronagraphic Imaging $5\sigma$ contrast curves at with F444W filter: \textit{ (left)} We show  contrast curves simulated using \texttt{PanCAKE} for three NIRCam exposure times in ADI and RDI mode.  We predict that if a cloudless exoplanet existed with a mass greater than 1\mjup outside of $\sim 1.5$\,AU, it would be detectable with 20 minutes of integration time.   \textit{(right)} The NIRCam F444W 5$\sigma$ ADI contrast curves were converted to mass space using the \citealt{Linder2019} models with an adopted age of 500 Myr. We find that exoplanets larger than 1 Saturn mass will be detectable outside of 2\,AU if Wolf 359 is in the younger part of its age range. A Neptune mass planet would be detectable beyond 6\,AU if Wolf 359 is younger than 500 Myr and a $>10$ hr exposure was used. }
\label{fig:PanCAKE}
\end{figure*}

Our simulated NIRCam contrast curves are shown in Figure \ref{fig:PanCAKE}.  Table \ref{tab:JWST} summarizes the detectability of theoretical cloudless exoplanets with varying masses using NIRCam in ADI mode with 1 hour of total integration time.  While our simulations span from 1-7\,AU (0.4\arcsec - 3\arcsec), the full NIRCam field of view from the MASK335R inner working angle (0.57\arcsec) to 20\arcsec would correspond to 1.4 - 48.2\,AU.  We estimate that the region from 7-48.2AU will be background limited and have the same contrast as the result at 7AU for future observing planning purposes.

One hour of NIRCam integration time would provide sensitivity to a cloudless Jupiter-mass companion outside of 0.62\arcsec (1.5\,AU) at any predicted age range. Cloudless Saturn-mass exoplanets (0.3 \mjup) would be detectable at small separations if Wolf 359 is in the youngest part of its age range and at wider background-limited separations for ages up to $\sim$1\,Gyr. A Neptune-like exoplanet (17 \mearth, 0.06 \mjup) will be visible if it is orbiting at wider separations and Wolf 359 is in the youngest part of its age range. 
The detection of a cloudless sub-Neptune exoplanet is unlikely with 1hr NIRCam ADI at any separations within Wolf 359's age range. 

\begin{deluxetable*}{c c c c}
\tablecaption{Summary of the NIRCam F444W Cornagraphic Imaging Detectability of Cloudless Companions.\label{tab:JWST}}
%\tabletypesize{\largesize}
\tablehead{\colhead{Planet Mass} & \colhead{Age} &\colhead{\begin{tabular}[c]{@{}c@{}}Predicted apparent \\ F444W mag\end{tabular}} &\colhead{\begin{tabular}[c]{@{}c@{}}Sep. where Detectable \\ by NIRCam 1 hr ADI \end{tabular}} }
\startdata
$1 M_{jup}$   & 100 Myr  & 12.90 & $>0.6 AU$      \\
$1 M_{jup}$   & 300 Myr  & 14.42 & $>0.8 AU$      \\
$1 M_{jup}$   & 500 Myr  & 15.19 & $>0.9 AU$      \\
$1 M_{jup}$   & 1200 Myr & 17.05 & $>1.4 AU$      \\
$1 M_{jup}$   & 1500 Myr & 17.34 & $>1.5 AU$      \\ \hline
$0.5 M_{jup}$ & 100 Myr  & 14.32 & $>0.8 AU$      \\
$0.5 M_{jup}$ & 300 Myr  & 16.09 & $>1.1 AU$      \\
$0.5 M_{jup}$ & 500 Myr  & 17.08 & $>1.4 AU$      \\
$0.5 M_{jup}$ & 1200 Myr & 19.51 & $>3.7 AU$      \\
$0.5 M_{jup}$ & 1500 Myr & 20.37 & $>4.7 AU$      \\ \hline
50 \mearth      & 100 Myr  & 17.06 & $>1.4 AU$      \\
50 \mearth      & 300 Myr  & 19.37 & $>3.6 AU$      \\ 
50 \mearth      & 500 Myr  & 20.85 & $>5.7 AU$      \\ \hline
20 \mearth      & 100 Myr  & 19.70 & $>3.9 AU$      \\
20 \mearth      & 300 Myr  & 23.33 & Not Detectable       
\enddata
\end{deluxetable*}

\subsubsection{MIRI Imaging}

\begin{figure}[ht!]
\plotone{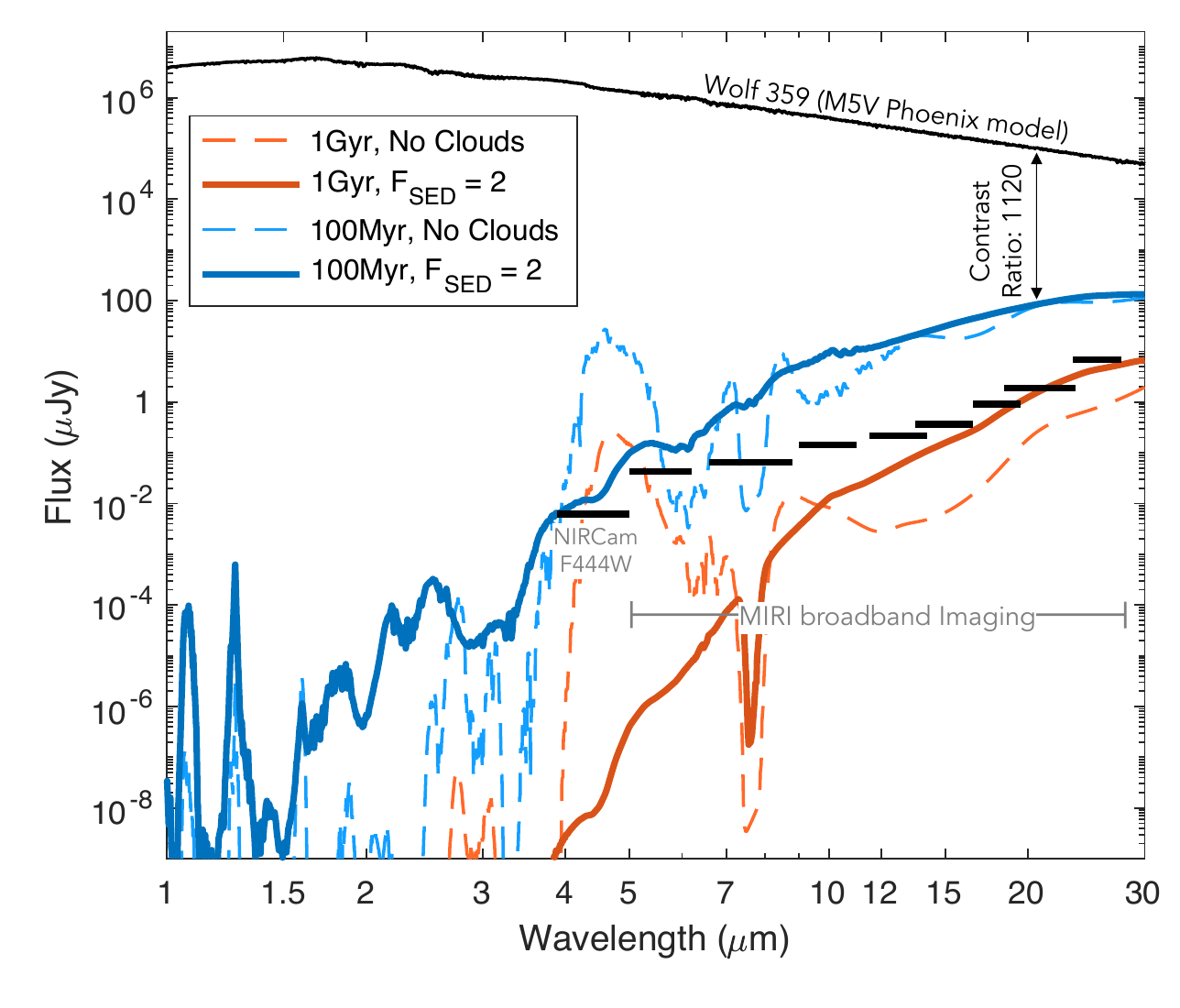}
\caption{Emission from a cloudy (solid lines) and clear (dashed lines) sub-Saturn (0.12\mjup) at 100\,Myr (blue) and 1\,Gyr (red): The Black line shows the emission from the star, Wolf 359 assuming a M5V spectral type. The black bars show the 3\,hr, 5$\sigma$ detection limits of NIRCam F444W and MIRI broadband imaging.  At 21$\mu m$, the contrast ratio between the star and a 100\,Myr, 0.12\mjup exoplanet is only 1120$\times$. For the older exoplanet, the contrast ratio is 15,800$\times$.}
\label{fig:clouds}
\end{figure}

Exoplanet gas giants with clear atmospheres are particularly bright in the emission band between 4-5$\mu m$, often making them detectable by the JWST NIRCam instrument. However, gas giants with cloudier atmospheres have muted emission from 4-5$\mu m$, instead emitting more at longer wavelengths ($>$15$\mu m$), as illustrated in Figure \ref{fig:clouds}. This figure shows the emission differences between a cloudy (solid lines) and clear (dashed lines) young, sub-Saturn exoplanet (0.12\mjup). The cloudy and clear models used in this figure were generated using the method described in \cite{Limbach2022}. This figure demonstrates that exoplanets with cloudy atmospheres may be more easily detected through JWST Mid-Infrared Instrument (MIRI) broadband imaging at 21$\mu m$, while clear atmospheres are more readily detected through direct imaging with NIRCam at 4.5$\mu m$.

We briefly explore the possibility of imaging exoplanets, like the Wolf 359b candidate from \citealt{Tuomi2019}, with MIRI.
In the mid-IR, the planet’s emission is increasing and the star’s emission is decreasing. This results in a favorable contrast ratio of the planet to star of 1:1120 for an exoplanet that is 100\,Myr, 0.12\mjup exoplanet with moderate cloud cover ($f_{SED}$ = 2). However, the diffraction limit of JWST at 21\,$\mu$m is 0.67\arcsec (6 pix) which is comparable to the separation between Wolf 359b and the host star. Using the coronagraphic mask at 23 $\mu$m, which has an inner working angle of 3.3$\lambda$/D, would block exoplanets at separations $<$2.16 arcsecs. Therefore, we instead consider directly imaging the system without a coronagraph and using KLIP \citep{Soummer2012} in post-processing to recover the exoplanet. KLIP has the potential to improve contrast by approximately $\sim100\times$ \citep{2015ApJ...809L..33R}.

\begin{figure}[ht!]
\plotone{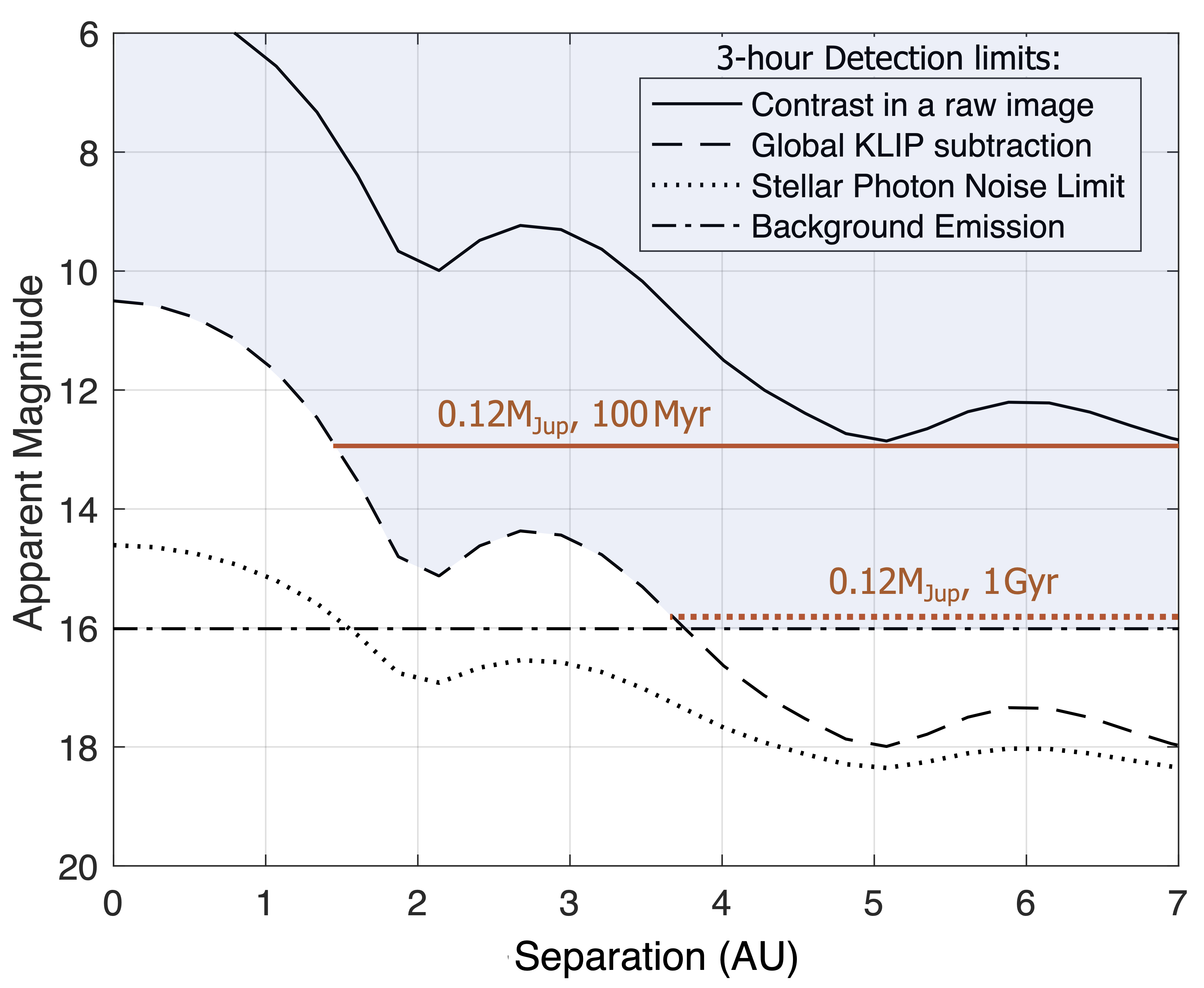}
\caption{Simulated contrast curve for JWST broadband imaging at 21$\mu m$. The solid black line shows the contrast in the raw image, the dashed black line is the residual after KLIP (assuming it is possible to achieve performance comparable to \citealt{2015ApJ...809L..33R}), and the dotted block line shows the photon-noise limit from stellar flux. The dash-dotted black line shows the 5\,$\sigma$ noise limit due to background emission. The shaded blue region above the KLIP contrast line and background emission line indicates the parameter space where it should be possible to detect exoplanets. The apparent magnitude of cloudy 100\,Myr and 1\,Gyr, 0.12\mjup exoplanets is shown by the red lines. This shows that a cloudy 100\,Myr, 0.12\mjup exoplanet should be detectable at separations $>$\,1.5\,AU, and a cloudy 1\,Gyr, 0.12\mjup exoplanet is detectable at separations $>$\,4\,AU. For this simulation, we assumed [F2100W] = 5.3\,mag, based on the star's WISE band 4 ($\lambda$=22.2\,$\mu$m) magnitude.}
\label{fig:21umContrast}
\end{figure}

Figure \ref{fig:21umContrast} shows the simulated MIRI contrast curve. To create this simulation, we used the pre-made set of point spread functions for JWST MIRI based on the in flight optical performance {\tt WebbPSF} tool\footnote{\url{jwst-docs.stsci.edu/jwst-mid-infrared-instrument/miri-performance/miri-point-spread-functions}}. We used the F21000W PSF that includes geometric optical distortions. The contrast curve for KLIP was calculated assuming performance similar to that described in \cite{2015ApJ...809L..33R}.

In Figure \ref{fig:21umContrast}, the shaded blue region above the black dashed line indicates the detectable exoplanet parameter space. With 3 hours of observation and using KLIP, a 0.12 \mjup planet with an age of 100\,Myr with moderate cloud cover would be detectable at separations greater than 1.5\,AU. With the same 3\,hr integration time, an older (1 Gyr) exoplanet of this size would also be detectable if at wider separations ($> 4$ AU). This approach requires an integration time which could fit into the JWST small proposals program and  
has the potential to detect nearby exoplanets to remarkably low masses.

\section{Conclusions} \label{sec:conclusions}

We conducted a joint high-contrast imaging survey and radial velocity survey with the goal of constraining long-period companions around the nearby M-dwarf star Wolf 359. We do not  rule out or confirm the Wolf 359\,b RV candidate as presented by \citealt{Tuomi2019}. 

To define the companion mass upper limits placed by our imaging  search, we performed an updated age analysis of Wolf 359 through kinematic age dating, CMD young moving group comparisons, and a MIST stellar isochrone comparison. We draw a conclusion of relative youth from the star's rotation period, and adopt the kinematic age of $1.53 \pm 0.3$\,Gyr as the upper bound for Wolf 359's age.  We rule out age estimates that are younger than $112$\,Myr through the comparison with young moving groups. Our MIST isochrone analysis produced an age estimate of $400$\,Myr.   

We conducted a high-contrast imaging survey using Keck-NIRC2 with the Ms filter (4.67\,$\mu m$) in conjunction with the vector vortex coronagraph. We totalled 4.98\,hr of integration time spread across 3 half-nights. The completeness of our imaging survey is highest (95\%) for the semi-major axis range from 1-3\,AU.  Our HCI results rule out a stellar or brown dwarf companion with this semi-major axis range to $5\sigma$, and 
companions smaller than $0.4$\mjup cannot be ruled out at any separation assuming an age older than $100$\,Myr. We compared our HCI survey's predicted performance as estimated by the NIRC2 SNR Calculator to our measured 5$\sigma$ performance and found a discrepancy of 1.7 magnitudes for the night of 2021 March 31 (UT).  This discrepancy can be partially accounted for by adjusting for the throughput loss when using the vortex at 4.7$\mu m$ and the fixhex pupil stop. Our analysis suggests that the remaining performance discrepancy may be due to the background noise exceeding the expected Poisson-noise level over time,  
indicating that it may be possible to improve the sensitivity of future surveys using faster image readout to better compensate for changes in the sky background.

We performed an updated radial velocity analyses of Wolf 359 with the \texttt{RVSearch} and \radvel python packages with data from four RV instruments: CARMENES, HARPS, \keckhires, and MAROON-X. 
After removing the known RV signal caused by the stellar rotation, we detect no signals above a false alarm probability of $0.1\%$. To 2$\sigma$, we exclude planets with a minimum mass bigger than \mpsini $> 13.5$ \mearth (0.0425 \mjup) with a semi-major axis smaller than $a < 0.1$ AU and planets with a minimum mass larger than \mpsini $> 147$ \mearth (0.46 \mjup) for a semi-major axis of less than $a<1$ AU.

We simulated JWST NIRCam and MIRI observations to explore the potential of JWST to directly image ice giant and gas giant exoplanets orbiting Wolf 359.  We predict that NIRCam Coronagraphic Imaging could detect a cloudless exoplanet with masses $>1$\mjup outside 1.5\,AU and $>0.5$\mjup outside 4.7\,AU with 1 hour of integration time (assuming an age younger than $<1.5$\,Gyr). Saturn and Neptune-mass exoplanets are accessible to NIRCam in certain  age/separation spaces, and it is unlikely that NIRCam could detect a sub-Neptune mass exoplanet. 
While MIRI imaging does not perform as well at smaller inner working angles, MIRI is capable of detecting cloudy exoplanets at smaller masses.  We predict that a cloudy companion with a mass of 0.12\mjup could be directly imaged to $5\sigma$ if orbiting outside 4\,AU using 3\,hours of integration time (assuming an age of younger than $1$\,Gyr).

This survey of Wolf 359 further establishes the methods needed to comprehensively characterize exoplanet systems using the intersection of multiple measurement techniques.  
As our future direct imaging instrumentation and RV surveys gain an increased sensitivity to ice giant exoplanets and super-Earths, the Wolf 359 system will continue to be a compelling target for understanding the cold planet population and planet formation outside the snow line of low-mass stars.

\section{Acknowledgements} \label{sec:acknowledgements}

The authors wish to recognize and acknowledge the very significant cultural role and reverence that the summit of Maunakea has always had within the indigenous Hawaiian community.  We are most fortunate to have the opportunity to conduct observations from this mountain.

RBR would like to thank Mikko Tuomi and Ignasi Ribas for their collaboration to include the HARPS-TERRA and CARMENES radial velocity data products. RBR also thanks Ester Linder, Jonathan Fortney, Andrew Skemer, Jorge Llop-Sayson, Andrew Howard, Caroline Morley, Kevin McKinnon, Kevin Wagner, Steve Ertl, Jason Wang, and Zack Breismeister  for lending their scientific expertise. RBR thanks Jules Fowler for their endless sound-boarding, python help, and title suggestion for this paper.

The authors would like to acknowledge the Keck staff who supported this observation including the observing assistants, Arina Rostopchina and Julie Renaud-Kim, and the instrument scientists, Carlos Alvarez and Greg Doppmann.  We thank Charlotte Bond and Sam Ragland who supported operation of the pyramid wavefront sensor and the following observers for their contribution in collecting the HIRES velocities: Isabel Angelo, Corey Beard, Aida Behmard, Sarah Blunt, Fei Dai, Paul Dalba, Benjamin	Fulton, Steven Giacalone, Rae Holcomb, Emma Louden, Jack Lubin, Andrew Mayo, Daria Pidhorodetska, Alex Polanski, Malena Rice, Emma	Turtelboom, Dakotah	Tyler, Lauren Weiss, and Judah Van Zandt.

The data presented were obtained at the W. M. Keck Observatory, which is operated as a scientific partnership among the California Institute of Technology, the University of California, and the National Aeronautics and Space Administration. The Observatory was made possible by the financial support of the W. M. Keck Foundation.

The University of Chicago group acknowledges funding for the MAROON-X project from the David and Lucile Packard Foundation, the Heising-Simons Foundation, the Gordon and Betty Moore Foundation, the Gemini Observatory, the NSF (award number 2108465), and NASA (grant number 80NSSC22K0117). We thank the staff of the Gemini Observatory for their assistance with the commissioning and operation of the instrument. The Gemini observations are associated with programs GN-2021A-Q-119, GN-2021B-Q-122, and GN-2022A-Q-119.

GS acknowledges support provided by NASA through the NASA Hubble Fellowship grant HST-HF2-51519.001-A awarded by the Space Telescope Science Institute, which is operated by the Association of Universities for Research in Astronomy, Inc., for NASA, under contract NAS5-26555.

J.M.A.M. is supported by the National Science Foundation (NSF) Graduate Research Fellowship Program under Grant No. DGE-1842400. J.M.A.M. acknowledges the LSSTC Data Science Fellowship Program, which is funded by LSSTC, NSF Cybertraining Grant No. 1829740, the Brinson Foundation, and the Moore Foundation; his participation in the program has benefited this work.

\software{QACTIS IDL software package \citep{Huby2017}, VIP: Vortex Imaging Processing python package \citep{VIP}, Species \citep{Stolker2020species}, Exo-DMC \citep{Bonavita2020}, RVSearch \citep{rosenthal21}, radvel \citep{radvel}, PanCAKE \citep{Girard2018, Perrin2018, CarterPancake}, galpy \citep{galpy},  Astropy \citep{astropy:2013, astropy:2018, astropy:2022}.}

\bibliography{sample631}{}
\bibliographystyle{aasjournal}

\clearpage
\appendix

\section{Supplemental Radial Velocity Information \label{sec:RVappend}}

A sample of the RV measurements used to complete the \texttt{RVSearch} analysis is listed in Table \ref{tab:rvs}.  
The full RV data set contains 275 velocities compiled from CARMENES, HARPS, \keckhires, and MAROON-X. 
The measurements made by our MAROON-X and \keckhires observations are available in Table \ref{tab:hiresrv} and Table \ref{tab:maroonxrv}.  These three tables are available in completion in machine readable format online.

\startlongtable
\begin{deluxetable*}{cccc}
\tablecaption{Velocities used in the  \texttt{RVSearch} analysis  \label{tab:rvs}}
\centering
\tablehead{
  \colhead{Time (BJD - 2400000)} & 
  \colhead{RV [m/s]} & 
  \colhead{RV Unc. [m/s]} & 
  \colhead{Inst.} 
}
\startdata
57397.72509 & -10.76 & 2.10 & CARMENES \\
57401.67629 & 6.00   & 1.18 & CARMENES \\
57419.56606 & -11.86 & 1.32 & CARMENES \\
57444.58536 & 3.66   & 1.52 & CARMENES \\
57449.637   & -1.18  & 1.61 & CARMENES\\
... & ... & ... & ... \\
59688.83007 & 1.37   & 1.07 & MAROONX$_\mathrm{red}$ \\
59689.92024 & -12.65 & 1.02 & MAROONX$_\mathrm{red}$ \\
59690.82035 & -3.38  & 1.03 & MAROONX$_\mathrm{red}$ \\
59695.91704 & -4.68  & 1.04 & MAROONX$_\mathrm{red}$ \\
59696.82571 & -1.51  & 1.03 & MAROONX$_\mathrm{red}$
\enddata
\tablecomments{The 5 first and 5 last velocities that were used in the \texttt{RVSearch} analysis are shown here as an example. HIRES$_\mathrm{k}$ and HIRES$_\mathrm{j}$ refer to \keckhires observations made before and after the detector upgrade in 2004.}
\end{deluxetable*}

\begin{longrotatetable}
\startlongtable
\begin{deluxetable*}{cccccccccccc}
\tablecaption{Radial velocity measurements for observations completed with \keckhires \label{tab:hiresrv}}
\centering
\tablehead{
  \colhead{Time (UT)} & 
  %\colhead{Telescope} & 
  \colhead{BJD} & 
  \colhead{RV [m/s]} &
  \colhead{RV Unc. [m/s]}&
\colhead{cts} & 
  \colhead{mdchi}& 
  \colhead{bc [m/s]} & 
  \colhead{svalue} & 
  \colhead{svalue$_{err}$} & 
  \colhead{trv [km/s]} & 
  \colhead{trv$_{err}$ [km/s]} 
  }
\startdata
2019-02-18 09:53:43.341   & 2458532.912307 & -28.3243 & 4.7225 & 1783.0 & 1.3819 & 7552.995   & 65.23 & 0.001 & 19.25   & 0.1 \\
2019-03-17 07:57:58.734   & 2458559.83193  & -20.5685 & 4.3440 & 1763.0 & 1.3092 & -6334.391  & 59.76 & 0.001 & 18.96   & 0.1 \\
2020-12-04 13:21:27.239   & 2459188.056565 & 6.8819   & 2.7510 & 5117.0 & 1.6743 & 30511.618  & 49.17 & 0.001 & 18.97   & 0.1 \\
2021-01-18 10:23:16.919   & 2459232.932835 & -4.1504  & 3.9815 & 4845.0 & 1.6342 & 21581.42   & 84.43 & 0.001 & 19.43   & 0.1 \\
2021-02-23 08:01:42.226   & 2459268.834516 & 10.9828  & 3.2371 & 4871.0 & 1.6621 & 4894.409   & 69.74 & 0.001 & 19.31   & 0.1 \\
2021-04-09 11:18:43.373   & 2459313.971335 & -5.0129  & 3.1419 & 5088.0 & 1.6624 & -17770.269 & 67.85 & 0.001 & -101.04 & 0.1 \\
2021-06-14 07:33:53.169   & 2459379.815199 & -10.2586 & 3.1191 & 4636.0 & 1.6224 & -29242.665 & 73.78 & 0.001 & 18.92   & 0.1 \\
2021-06-24 07:15:16.741   & 2459389.802277 & 5.8263   & 3.1427 & 4664.0 & 1.6652 & -27939.286 & 108.0 & 0.001 & 18.68   & 0.1 \\
2021-12-17 12:05:46.440   & 2459566.00401  & 3.7792   & 2.8511 & 4541.0 & 1.6260 & 29791.676  & 61.49 & 0.001 & 19.2    & 0.1 \\
2022-01-13 10:21:53.730   & 2459592.931872 & -9.8209  & 3.8080 & 3774.0 & 1.5332 & 23485.419  & 70.49 & 0.001 & 19.17   & 0.1 \\
2022-02-21 08:31:28.284   & 2459631.855188 & -11.4183 & 3.3142 & 4584.0 & 1.5754 & 6033.566   & 78.8  & 0.001 & 19.02   & 0.1 \\
2022-02-22 09:08:41.970  & 2459632.881041 & -0.4443  & 3.4243 & 4284.0 & 1.6415 & 5444.217   & 82.33 & 0.001 & 19.08   & 0.1
\enddata
\tablecomments{\small cts - Counts in raw 1D spectrum near 5500 Angstroms [e-], \\
chi - Median reduced chi-squared for the observation over all chunks, \\
bc - Barycentric velocity at flux-weighted midpoint [m/s], \\
svalue - CaHK S-value, \\
svalue$_{err}$ - CaHK S-value uncertainty, \\
trv - Telluric calibrated absolute RV [km/s], \\
trv$_{err}$ - Uncertainty in telluric-calibrated absolute RV [km/s].  }
\end{deluxetable*}
\end{longrotatetable}

\begin{longrotatetable}
\startlongtable
\begin{deluxetable*}{ccccccccccccccccc}
\tablecaption{Radial velocity measurements for observations completed with MAROON-X \label{tab:maroonxrv}}
\centering
\tablehead{
 \colhead{Arm}& 
 \colhead{BJD} &
 \colhead{RV\_po} & 
 \colhead{e\_RV\_po} &
 \colhead{snpeak} & 
 \colhead{exptime} &
 \colhead{berv }& 
 \colhead{airmass }&
 \colhead{dLW} & 
 \colhead{e\_dLW} & 
 \colhead{crx} & 
 \colhead{e\_crx } & 
 \colhead{off. epoch} & 
 \colhead{offset} & 
 \colhead{e\_offset }& 
 \colhead{RV}&
 \colhead{e\_rv}
   }
\startdata
blue & 2459267.8235215 & -14.64 & 1.35 & 66.0  & 1800 & 5.43   & 1.72 & -7.79  & 1.87 & -5.21  & 23.03 & 1 & -1.5 & 0.5 & -13.14 & 1.44 \\
blue & 2459269.9620281 & -8.71  & 0.94 & 81.0  & 1800 & 3.97   & 1.03 & -16.21 & 1.29 & -9.95  & 15.60 & 1 & -1.5 & 0.5 & -7.21  & 1.07 \\
blue & 2459320.8841243 & 3.48   & 0.82 & 94.0  & 1800 & -20.33 & 1.09 & 8.75   & 1.11 & -1.58  & 11.83 & 2 & 0.0  & 0.0 & 3.48   & 0.82 \\
blue & 2459321.8689137 & -11.94 & 1.05 & 72.0  & 1800 & -20.66 & 1.06 & -30.50 & 1.45 & -4.65  & 18.10 & 2 & 0.0  & 0.0 & -11.94 & 1.05 \\
blue & 2459322.9199499 & -0.89  & 0.91 & 88.0  & 1800 & -21.16 & 1.23 & -29.97 & 1.26 & 2.13   & 15.72 & 2 & 0.0  & 0.0 & -0.89  & 0.91 \\
blue & 2459332.9311944 & -6.96  & 1.27 & 71.0  & 1800 & -24.44 & 1.51 & -1.17  & 1.75 & 31.35  & 18.35 & 2 & 0.0  & 0.0 & -6.96  & 1.27 \\
blue & 2459333.8404301 & -0.58  & 0.68 & 109.0 & 1800 & -24.50 & 1.07 & -34.30 & 0.93 & 23.86  & 10.83 & 2 & 0.0  & 0.0 & -0.58  & 0.68 \\
blue & 2459334.8384658 & -10.13 & 0.68 & 108.0 & 1800 & -24.78 & 1.07 & -20.69 & 0.94 & 15.79  & 11.59 & 2 & 0.0  & 0.0 & -10.13 & 0.68 \\
blue & 2459359.8578599 & -9.48  & 1.45 & 64.0  & 1800 & -29.37 & 1.51 & 22.24  & 1.99 & -14.84 & 22.69 & 3 & 2.5  & 1.5 & -11.98 & 2.09 \\
blue & 2459361.8234208 & -12.07 & 0.86 & 99.0  & 1800 & -29.45 & 1.29 & -0.19  & 1.18 & -12.82 & 13.34 & 3 & 2.5  & 1.5 & -14.57 & 1.73 \\
blue & 2459362.8356031 & -3.80  & 1.17 & 82.0  & 1800 & -29.53 & 1.39 & 5.57   & 1.61 & -2.97  & 21.52 & 3 & 2.5  & 1.5 & -6.30  & 1.90 \\
blue & 2459363.8070943 & 5.54   & 0.84 & 101.0 & 1800 & -29.52 & 1.23 & 9.00   & 1.15 & 45.59  & 12.53 & 3 & 2.5  & 1.5 & 3.04   & 1.72 \\
blue & 2459364.8529316 & -14.66 & 1.25 & 81.0  & 1800 & -29.66 & 1.61 & 1.79   & 1.73 & -33.41 & 37.31 & 3 & 2.5  & 1.5 & -17.16 & 1.95 \\
blue & 2459367.7719152 & -9.82  & 0.88 & 94.0  & 1800 & -29.58 & 1.13 & -24.13 & 1.21 & -8.87  & 9.92  & 3 & 2.5  & 1.5 & -12.32 & 1.74 \\
blue & 2459368.8130652 & -1.22  & 0.89 & 97.0  & 1800 & -29.68 & 1.34 & -14.78 & 1.23 & 10.27  & 10.50 & 3 & 2.5  & 1.5 & -3.72  & 1.74 \\
blue & 2459515.1284389 & 7.45   & 1.54 & 66.0  & 1800 & 24.07  & 2.03 & 14.83  & 2.13 & -26.51 & 27.23 & 4 & 8.0  & 1.0 & -0.55  & 1.83 \\
blue & 2459518.140394  & 8.22   & 1.24 & 71.0  & 1800 & 24.98  & 1.69 & 27.88  & 1.70 & -17.33 & 17.08 & 4 & 8.0  & 1.0 & 0.22   & 1.59 \\
blue & 2459525.1065205 & -5.89  & 1.60 & 58.0  & 1800 & 26.94  & 1.92 & 15.17  & 2.21 & -8.92  & 27.06 & 4 & 8.0  & 1.0 & -13.89 & 1.89 \\
blue & 2459527.1188207 & 4.16   & 1.03 & 86.0  & 1800 & 27.40  & 1.65 & 8.78   & 1.42 & 10.32  & 14.81 & 4 & 8.0  & 1.0 & -3.84  & 1.44 \\
blue & 2459530.1308991 & 4.48   & 0.81 & 104.0 & 1800 & 28.02  & 1.44 & -1.47  & 1.12 & 13.60  & 12.24 & 4 & 8.0  & 1.0 & -3.52  & 1.29 \\
blue & 2459539.1271668 & 0.46   & 0.95 & 84.0  & 1800 & 29.49  & 1.28 & -1.42  & 1.31 & 2.73   & 9.44  & 4 & 8.0  & 1.0 & -7.54  & 1.38 \\
blue & 2459540.1210453 & 1.26   & 0.77 & 108.0 & 1800 & 29.62  & 1.31 & -8.13  & 1.05 & -0.08  & 9.74  & 4 & 8.0  & 1.0 & -6.74  & 1.26 \\
blue & 2459541.1118372 & -2.48  & 0.79 & 105.0 & 1800 & 29.75  & 1.35 & 1.00   & 1.08 & -9.11  & 10.39 & 4 & 8.0  & 1.0 & -10.48 & 1.27 \\
blue & 2459666.8346899 & 5.59   & 0.62 & 118.0 & 1800 & -11.91 & 1.07 & -12.53 & 0.84 & 12.81  & 7.19  & 5 & 11.0 & 1.0 & -5.41  & 1.17 \\
blue & 2459677.8746489 & 4.58   & 0.87 & 85.0  & 1800 & -17.01 & 1.04 & -11.46 & 1.19 & 3.43   & 10.10 & 5 & 11.0 & 1.0 & -6.42  & 1.32 \\
blue & 2459678.8934433 & -1.64  & 1.03 & 72.0  & 1800 & -17.49 & 1.06 & -7.99  & 1.42 & -1.25  & 12.33 & 5 & 11.0 & 1.0 & -12.64 & 1.43 \\
blue & 2459680.8488688 & 11.42  & 0.98 & 75.0  & 1800 & -18.17 & 1.03 & 13.24  & 1.34 & 13.21  & 12.82 & 5 & 11.0 & 1.0 & 0.42   & 1.40 \\
blue & 2459683.8770181 & 9.33   & 0.63 & 116.0 & 1800 & -19.44 & 1.06 & 42.70  & 0.84 & 13.57  & 7.16  & 5 & 11.0 & 1.0 & -1.67  & 1.18 \\
blue & 2459684.9261998 & -1.01  & 0.61 & 113.0 & 1800 & -19.96 & 1.22 & 10.71  & 0.83 & -4.58  & 7.21  & 5 & 11.0 & 1.0 & -12.01 & 1.17 \\
blue & 2459688.830072  & 11.27  & 1.59 & 48.0  & 1800 & -21.16 & 1.03 & 18.41  & 2.17 & 13.31  & 19.52 & 5 & 11.0 & 1.0 & 0.27   & 1.88 \\
blue & 2459689.9202404 & -5.46  & 0.57 & 121.0 & 1800 & -21.77 & 1.26 & -11.84 & 0.78 & -8.66  & 7.88  & 5 & 11.0 & 1.0 & -16.46 & 1.15 \\
blue & 2459690.820352  & 6.56   & 0.63 & 111.0 & 1800 & -21.84 & 1.03 & -21.48 & 0.87 & 12.18  & 8.73  & 5 & 11.0 & 1.0 & -4.44  & 1.18 \\
blue & 2459695.9170363 & 7.44   & 0.94 & 81.0  & 1800 & -23.75 & 1.35 & 17.30  & 1.28 & 8.29   & 13.65 & 5 & 11.0 & 1.0 & -3.56  & 1.37 \\
blue & 2459696.8257078 & 6.11   & 0.98 & 110.0 & 1800 & -23.82 & 1.04 & 127.82 & 1.22 & 25.24  & 12.37 & 5 & 11.0 & 1.0 & -4.89  & 1.40 \\
red  & 2459267.8235215 & -16.62 & 0.39 & 375.0 & 1800 & 5.43   & 1.72 & 60.96  & 0.45 & 29.81  & 7.62  & 1 & -1.6 & 0.5 & -15.02 & 0.63 \\
red  & 2459269.9620281 & -9.27  & 0.38 & 403.0 & 1800 & 3.97   & 1.03 & 65.80  & 0.43 & 0.20   & 6.09  & 1 & -1.6 & 0.5 & -7.67  & 0.63 \\
red  & 2459320.8841243 & -0.58  & 0.34 & 468.0 & 1800 & -20.33 & 1.09 & 71.81  & 0.37 & -22.98 & 4.82  & 2 & 0.0  & 0.0 & -0.58  & 0.34 \\
red  & 2459321.8689137 & -10.45 & 0.34 & 365.0 & 1800 & -20.66 & 1.06 & 55.56  & 0.41 & 13.75  & 6.07  & 2 & 0.0  & 0.0 & -10.45 & 0.34 \\
red  & 2459322.9199499 & -1.93  & 0.30 & 461.0 & 1800 & -21.16 & 1.23 & 55.41  & 0.34 & 0.01   & 3.79  & 2 & 0.0  & 0.0 & -1.93  & 0.30 \\
red  & 2459332.9311944 & -4.96  & 0.32 & 419.0 & 1800 & -24.44 & 1.51 & 54.08  & 0.38 & 12.32  & 4.48  & 2 & 0.0  & 0.0 & -4.96  & 0.32 \\
red  & 2459333.8404301 & 0.65   & 0.30 & 531.0 & 1800 & -24.50 & 1.07 & 55.95  & 0.34 & -5.79  & 3.50  & 2 & 0.0  & 0.0 & 0.65   & 0.30 \\
red  & 2459334.8384658 & -11.28 & 0.31 & 529.0 & 1800 & -24.78 & 1.07 & 63.81  & 0.34 & 25.58  & 6.89  & 2 & 0.0  & 0.0 & -11.28 & 0.31 \\
red  & 2459359.8578599 & -4.78  & 0.28 & 379.0 & 1800 & -29.37 & 1.51 & -21.86 & 0.38 & 10.07  & 4.68  & 3 & 2.2  & 1.0 & -6.98  & 1.04 \\
red  & 2459361.8234208 & -9.22  & 0.23 & 519.0 & 1800 & -29.45 & 1.29 & -18.08 & 0.31 & 14.87  & 4.46  & 3 & 2.2  & 1.0 & -11.42 & 1.03 \\
red  & 2459362.8356031 & -2.32  & 0.23 & 483.0 & 1800 & -29.53 & 1.39 & -22.59 & 0.31 & 0.39   & 3.41  & 3 & 2.2  & 1.0 & -4.52  & 1.03 \\
red  & 2459363.8070943 & 4.11   & 0.23 & 523.0 & 1800 & -29.52 & 1.23 & -9.55  & 0.31 & -33.92 & 5.02  & 3 & 2.2  & 1.0 & 1.91   & 1.03 \\
red  & 2459364.8529316 & -10.04 & 0.22 & 539.0 & 1800 & -29.66 & 1.61 & -22.92 & 0.29 & 19.62  & 5.67  & 3 & 2.2  & 1.0 & -12.24 & 1.02 \\
red  & 2459367.7719152 & -7.97  & 0.23 & 487.0 & 1800 & -29.58 & 1.13 & -25.37 & 0.30 & 9.56   & 4.71  & 3 & 2.2  & 1.0 & -10.17 & 1.03 \\
red  & 2459368.8130652 & 2.06   & 0.23 & 519.0 & 1800 & -29.68 & 1.34 & -20.19 & 0.30 & -15.18 & 4.62  & 3 & 2.2  & 1.0 & -0.14  & 1.03 \\
red  & 2459515.1284389 & 4.52   & 0.30 & 417.0 & 1800 & 24.07  & 2.03 & -24.05 & 0.40 & -11.83 & 5.67  & 4 & 7.0  & 1.0 & -2.48  & 1.04 \\
red  & 2459518.140394  & 3.98   & 0.30 & 389.0 & 1800 & 24.98  & 1.69 & -17.33 & 0.41 & -18.53 & 5.24  & 4 & 7.0  & 1.0 & -3.02  & 1.04 \\
red  & 2459525.1065205 & -4.96  & 0.33 & 334.0 & 1800 & 26.94  & 1.92 & -19.60 & 0.46 & 22.67  & 7.77  & 4 & 7.0  & 1.0 & -11.96 & 1.05 \\
red  & 2459527.1188207 & 4.37   & 0.26 & 470.0 & 1800 & 27.40  & 1.65 & -19.71 & 0.35 & -10.88 & 4.60  & 4 & 7.0  & 1.0 & -2.63  & 1.03 \\
red  & 2459530.1308991 & 2.87   & 0.24 & 537.0 & 1800 & 28.02  & 1.44 & -21.33 & 0.32 & -9.00  & 3.84  & 4 & 7.0  & 1.0 & -4.13  & 1.03 \\
red  & 2459539.1271668 & 0.86   & 0.29 & 427.0 & 1800 & 29.49  & 1.28 & -27.30 & 0.38 & 12.48  & 4.96  & 4 & 7.0  & 1.0 & -6.14  & 1.04 \\
red  & 2459540.1210453 & 2.55   & 0.25 & 543.0 & 1800 & 29.62  & 1.31 & -26.80 & 0.32 & 4.92   & 3.75  & 4 & 7.0  & 1.0 & -4.45  & 1.03 \\
red  & 2459541.1118372 & 0.06   & 0.24 & 531.0 & 1800 & 29.75  & 1.35 & -20.85 & 0.32 & 9.26   & 3.13  & 4 & 7.0  & 1.0 & -6.94  & 1.03 \\
red  & 2459666.8346899 & 8.39   & 0.23 & 565.0 & 1800 & -11.91 & 1.07 & -28.82 & 0.29 & 0.66   & 3.68  & 5 & 10.0 & 1.0 & -1.61  & 1.03 \\
red  & 2459677.8746489 & 6.42   & 0.27 & 416.0 & 1800 & -17.01 & 1.04 & -30.48 & 0.35 & -6.16  & 4.43  & 5 & 10.0 & 1.0 & -3.58  & 1.04 \\
red  & 2459678.8934433 & 0.88   & 0.29 & 362.0 & 1800 & -17.49 & 1.06 & -26.49 & 0.38 & 7.39   & 5.50  & 5 & 10.0 & 1.0 & -9.12  & 1.04 \\
red  & 2459680.8488688 & 9.62   & 0.28 & 374.0 & 1800 & -18.17 & 1.03 & -24.81 & 0.38 & -21.12 & 4.33  & 5 & 10.0 & 1.0 & -0.38  & 1.04 \\
red  & 2459683.8770181 & 8.59   & 0.21 & 550.0 & 1800 & -19.44 & 1.06 & -19.94 & 0.27 & -14.60 & 3.76  & 5 & 10.0 & 1.0 & -1.41  & 1.02 \\
red  & 2459684.9261998 & 1.28   & 0.23 & 539.0 & 1800 & -19.96 & 1.22 & -26.99 & 0.29 & 5.97   & 3.37  & 5 & 10.0 & 1.0 & -8.72  & 1.03 \\
red  & 2459688.830072  & 11.37  & 0.39 & 242.0 & 1800 & -21.16 & 1.03 & -27.16 & 0.53 & -19.19 & 5.03  & 5 & 10.0 & 1.0 & 1.37   & 1.07 \\
red  & 2459689.9202404 & -2.65  & 0.22 & 580.0 & 1800 & -21.77 & 1.26 & -27.18 & 0.28 & 25.49  & 4.46  & 5 & 10.0 & 1.0 & -12.65 & 1.02 \\
red  & 2459690.820352  & 6.62   & 0.24 & 527.0 & 1800 & -21.84 & 1.03 & -29.38 & 0.30 & -7.89  & 3.67  & 5 & 10.0 & 1.0 & -3.38  & 1.03 \\
red  & 2459695.9170363 & 5.32   & 0.28 & 399.0 & 1800 & -23.75 & 1.35 & -25.53 & 0.37 & 0.19   & 3.56  & 5 & 10.0 & 1.0 & -4.68  & 1.04 \\
red  & 2459696.8257078 & 8.49   & 0.24 & 515.0 & 1800 & -23.82 & 1.04 & -11.83 & 0.33 & -10.49 & 5.78  & 5 & 10.0 & 1.0 & -1.51  & 1.03
\enddata
\tablecomments{\small bjd - Julian Date [days],\\
rv\_po/e\_rv\_po - Barycentric-corrected radial velocity (and its error) pre offset correction by observing run epoch [m/s],\\
sn\_peak - Peak signal-to-noise in observation,\\
exptime - exposure time [s]\\
berv - flux-weighted barycentric correction included in final velocities, standard barycentric correction to radial velocity [m/s],\\
airmass - observation airmass,\\
dLW/e\_dLW - Differential linewidth of observation (an activity indicator described in the serval documentation) [1000$m^2/s^2$],\\
crx/e\_crx - Chromatic index of observation (an activity indicator described in the serval documentation) [m/s/Np],\\
irt\_ind/e\_irt\_ind - CaIRT indices of observation,\\
off. epoch - Observing run epoch with an applicable instrument offset [m/s],\\
offset/e\_offset - radial velocity offset measured between observing runs [m/s],\\
rv/e\_rv - Barycentric-corrected and offset-corrected by observing epoch radial velocity (and its error) [m/s] }
\end{deluxetable*}
\end{longrotatetable}

\section{Revised SNR Estimations for Keck-NIRC2 with the Vortex Coronagraph in M-band \label{sec:revisedcalc}}

The W.M. Keck Observatory provides an online tool to help observers plan their NIRC2 observations, the NIRC2 SNR and Efficiency Calculator.\footnote{\url{https://www2.keck.hawaii.edu/inst/nirc2/nirc2_snr_eff.html}}   
The NIRC2 Calculator does not provide performance estimates for using the vector vortex coronagraph mask, which is a common configuration for conducting high-contrast imaging observations to hunt for exoplanet and brown dwarf companions. 
In this section, we offer a method to modify the equations used by the NIRC2 Calculator to aid in the SNR prediction when using the L/M-band vortex \citep{Serabyn2017} with the Ms filter.

We recommend that observers planning to use the Lp filter in conjunction with the L/M-band vortex use the Vortex Imaging Contrast Oracle (VICO)\footnote{\url{https://wxuan.shinyapps.io/contrast-oracle/}} \citep{Xuan2018} instead of the equations documented here. VICO produces a full contrast curve based off of user inputs for the host's magnitude, the survey's total integration time, and predicted spanned parallactic angle.  VICO's performance and contrast predictions are based off a training set of 304 targets that were observed between 2015 to 2018 using the Shack-Hartmann wavefront sensor to perform adaptive optics correction.   

We consider SNR predictions only and do not attempt to match the efficiency predictions made by the NIRC2 Calculator.
The predictions offered by our modified equations are intended to predict the SNR in the background limited regions where light from the host star is negligible in comparison to the background flux.  We do not attempt to predict the full contrast curve or the SNR in the speckle limited regions of the image. 

Table \ref{tab:nirc2snrcalc} shows the values defined internally to the NIRC2 SNR Calculator alongside our choices for the user defined parameters that are applicable to our observing mode.  
The calculator internally defines the read noise, gain, zero-point for each filter, and number of pixels within a full-width-half-max for each filter. The user is able to specify the object magnitude, Strehl ratio, time per exposure (tint), coadds, number of dithers, repeats per dither, camera mode (narrow versus wide), filter, number of reads, array window size, and adaptive optics mode (natural guide star versus laser guide star).   We matched our user defined parameters to our 2021 March 31 (UT) dataset to compare the predicted performance to the measured performance on that night.  Terms related to laser motion control were excluded because our observations were performed using natural guide star adaptive optics. 
We adopted a Strehl ratio of 0.85 for our predictions, which is a conservative estimate equivalent to 300\,nm of error on the wavefront.  

%Our modified equations are listed at the end of this section. 
We began the modifications to the equations used in the NIRC2 SNR Calculator by adding a corrective factor for the throughput for using Keck-II's fixedhex pupil stop.  The NIRC2 SNR calculator assumes that NIRC2 images will be taken with Keck Observatory's circular largehex pupil stop. However, when operating NIRC2 in conjunction with the vortex coronagraph, the fixedhex pupil is typically used.  The fixedhex pupil was specifically tailored for use with the vortex coronagraph at Keck, and its shape blocks the telescopes spiders and central obscuration. It has a throughput of 84\% as compared to the largehex pupil. We assigned a 0.84 throughput penalty for this difference in pupil stop. 

We further refined our throughput penalty by accounting for the throughput hit due to the absorption by the vector vortex coronagraph at 4.67$\mu m$. The throughput of the Keck annular groove phase mask was measured in the lab to be $70 \pm 3\%$ (\citealt{Jolivet2019}, AGPM-L9r2 in Table 4).
We conducted on-sky testing to verify this throughput as observed with the full optical system in June 2022.  We imaged HIP 74785 using the fixhexed pupil for all images. We moved the vortex coronagraph in-and-out of the optical path to create a direct comparison to measure the transmission of the vortex mask. When the vortex was in place, it was intentionally miscentered with respect to the stellar PSF to assure no flux from the star was blocked due to the coronagraphy properties of the vortex.  The stellar photometry was then measured in each image using \texttt{photutils} within a circular aperture of 60\,pixels in radius.  The throughput ratio of the no-vortex to with-vortex stellar flux was measured to be $68 \pm 3\%$.  While the vortex throughput measurement using on-sky images is consistent with laboratory tests, we note that we find that there was a slight defocus in the on-sky images when the vortex was in place such that a photometry aperture radius of $>5\,FWHM$ needed to be used to achieve this consistency. 

  We document both the lab and on-sky testing values for the throughput penalty due to the vortex absorption in Table \ref{tab:nirc2snrcalc} alongside our calculation for the total throughput penalty. The total throughput penalty was calculated by multiplying the vortex throughput penalty by the fixhex throughput penalty.   We adopted the more pessimistic value of the total throughput penalty (57\%) when discussing the performance of NIRC2 in Section \ref{sec:discussion}.

To quantify a correction factor for the background flux, we measured the background flux per pixel in each image for our full datacube from the three observing nights of Wolf 359. 
 The NIRC2 calculator assumes a value of 18535\,DN/s per pixel for the sky background contribution when observing in narrow mode with the Ms filter using no vortex with the largehex pupil. 
We measured the sky background in our images with the vortex and fixhex pupil by finding the median of each image cube. We then averaged those medians by night to find the average image median to be 17457 $\pm$ 4 on 2021 Feb 22,
17008 $\pm$ 7 DN/s on 2021 Feb 23, and
17850 $\pm$ 106 DN/s on 2021 March 31. These values indicate that there is a measurable excess background flux when the vortex optic is in place, as the background counts do not scale with the throughput penalty.  We adopt the most pessimistic value for the background flux (2021 March 31) for our discussion of the NIRC2 performance in Section \ref{sec:discussion}. 

The equations outlined in this appendix account for the throughput penalty of using the vortex only.
When applied to the Wolf 359 2021 March 31 dataset, this method of prediction overestimated the $5\sigma$ performance capabilities by $\sim1$ magnitude.  We caution observers who are planning to detect a companion dimmer than $m_s = 14.4$ with Keck-NIRC2 using one or more half-nights approach the observation carefully and not rely solely on the online NIRC SNR calculator or the equations listed in this section.

\begin{deluxetable*}{l c c l}
\tablecaption{Parameters used in the modified NIRC2 Performance Calculation }
\label{tab:nirc2snrcalc}  
%\tabletypesize{\scriptsize}
\tablehead{ \colhead{Parameter} & \colhead{Value} & \colhead{How defined?} & \colhead{Notes}  }
\startdata
Read noise in electrons ($r_{noise}$) & 56 & Internal  &  \\ \hline
Gain in e- per DN ($gain$)  & 4 & Internal  &   \\ \hline
\begin{tabular}[c]{@{}l@{}}The zero-point for Ms filter \\ in narrow mode ($zeropoint$)\end{tabular}    & 22.7 & Internal  &  \\ \hline
\begin{tabular}[c]{@{}l@{}}Number of pixels in \\ aperture ($n_{pix}$)\end{tabular}   & 490.8  & Internal  &  \\ \hline \hline
N Reads  & 2 (CDS)  & User  & 2 reads is a standard readout mode  \\ \hline
Array Window Size      & 512  & User & 512 x 512 is a standard img size for HCI imaging    \\ \hline
AO Mode    & NGS    & User  & Natural Guide star mode   \\ \hline
\textit{Strehl} & 0.85  & User   & Conservative assumption assuming 300\,nm of wavefront error\\ \hline
\textit{tint}     & 0.3\,s   & User & Time per frame before coadd used in this survey   \\ \hline
\textit{coadds}    & 90  & User  & Num. frames added together before a full read used in this survey \\ \hline
Number of images ($n_{exp}$)   & 269   & User     & Num. images collected on 2021 Mar 31 of this survey               \\ \hline \hline
%\textit{Fixhex Throughput}  
\begin{tabular}[c]{@{}l@{}}\textit{Fixhex Throughput} as compared \\ to largehex pupil \end{tabular} 
& 0.84    & Calculated  & The ratio of the size of the fixedhex/largehex pupil at Keck      \\ \hline
\begin{tabular}[c]{@{}l@{}}\textit{Vortex Throughput} at 4.6\,$\mu$m \\ meas. in lab\end{tabular}                         & 0.70 $\pm$ 0.03 & Measured   & AGPM-L9r2 in Table 4 of \citealt{Jolivet2019}    \\ \hline
\begin{tabular}[c]{@{}l@{}}\textit{Vortex Throughput} at 4.6\,$\mu$m \\ meas. on sky\end{tabular}                      & $0.68 \pm 0.03 $           & Measured          & Meas. using data shared by Keck Obs staff                                           \\ \hline
\begin{tabular}[c]{@{}l@{}}\textit{Throughput Penalty} using \\ vortex value meas. in lab\end{tabular} & $0.59 \pm 0.03$ & Calculated    & $FixhexThrouph*VortexThrouput_{lab}$      \\ \hline
\begin{tabular}[c]{@{}l@{}}\textit{Throughput Penalty} using \\ vortex value meas. on sky\end{tabular}   & $0.57 \pm 0.03$  & Calculated   & \begin{tabular}[c]{@{}l@{}}$FixhexThrouph*VortexThrouput_{sky}$; Value adopted\\ for NIRC2 performance discussion in Section 4. \end{tabular}    \\ \hline
\begin{tabular}[c]{@{}l@{}}\textit{Background} flux using vortex \\  with fixhex pupil at 4.6\,$\mu$m\end{tabular} & 17850\,DN/s     & Measured     & Average DN/s meas. on 2021 Mar 31 of this survey                 
\enddata
%\tablenotetext{}{text}
\end{deluxetable*}

\textbf{Equations to calculate the SNR:}

The terms that deviate from the formulas used by the NIRC2 SNR calculator are highlighted in red.

\begin{equation*}
\begin{aligned}
n_{exp} &= coadds*n_{images} & \text{We assume no dithering} \\
m_{zero} &= zeropoint + 2.5 * \log_{10}(strehl) &\text{} \\
bg &=  {\color{red} background} * gain  & \text{Meas. $background$ = 17850 DN/s per pix} \\
\\
 ThroughputPenalty &= {\color{red} FixhexThroughput * VortexThroughput } &\text{Meas. $ThroughputPenalty$ = 0.58}\\
%&=0.46 ~or~ 0.59 &\text{We adopt 0.46 for our further evaluations}
\\
signal &= {\color{red} ThroughputPenalty} * n_{exp} * tint * 10^{(0.4 (m_{zero} - mag))}&\text{Signal after throughput penalty applied} \\
\\
\sigma_{read}^2 &= n_{exp} * r_{noise}^2 * n_{pix} &\text{Read noise}\\
\sigma_{skybg}^2 &= n_{pix} * bg * n_{exp} * tint &\text{Photon-noise of sky background}\\
\sigma_{photon}^2 &= signal &\text{Photon-noise of source}\\
\sigma_{tot} &= \sqrt{\sigma_{read}^2 + \sigma_{skybg}^2 + \sigma_{photon}^2} &\text{Total noise}\\
\\
SNR &= signal / \sigma_{tot} &\text{Signal-to-Noise}
\end{aligned}
\end{equation*}

\end{document}